\documentstyle[12pt]{article}
\input epsf
\begin{document}
\renewcommand{\thefootnote}{\fnsymbol{footnote}}
%%%%%%%%%%%%%%%%%%%%%% title page %%%%%%%%%%%%%%%%%%%%%%%%%%%%%%
\begin{titlepage}
\renewcommand{\thefootnote}{\fnsymbol{footnote}}
\makebox[2cm]{}\\[-1in]
\begin{flushright}
\begin{tabular}{l}
DESY 94-181\\
hep-ph/9410318
\end{tabular}
\end{flushright}
\vskip0.4cm
\begin{center}
{\Large\bf
Ioffe-time distributions instead of parton momentum distributions
     in description of deep inelastic scattering             }

\vspace{2cm}

V.\ Braun$^1$\footnote { On leave of absence from
St.Petersburg Nuclear
Physics Institute, 188350 Gatchina, Russia},
P.\ G\'ornicki$^2$ and L.\ Mankiewicz$^3$

\vspace{1.5cm}

$^1${\em DESY, Notkestra\ss\/e 85, D--22603 Hamburg, Germany}\\[0.5cm]
$^2${\em     Center for Theoretical Physics and Institute of Physics,
    Polish Academy of Sciences,
    Al. Lotnikow 32/46, PL--02--668 Warsaw (Poland)  }\\[0.5cm]
$^3${\em N. Copernicus Astronomical Center,
        Polish Academy of Science,
        ul. Bartycka 18, PL--00-716 Warsaw (Poland)  }\\[0.5cm]

\vspace{1cm}

{\em \today}

\vspace{1cm}

{\bf Abstract:\\[5pt]}
\parbox[t]{\textwidth}{
We argue that parton distributions in coordinate space provide a more
natural object for nonperturbative methods compared to the usual
momentum distributions in which the physics of different longitudinal
distances is being mixed. To illustrate the advantages of the
coordinate space formulation, we calculate the coordinate space
distributions for valence quarks in the proton using the QCD sum rule
approach.  A remarkable agreement is found between the calculated and
the experimentally measured u-quark distribution up to light-cone
distances $\Delta^- = \Delta^0 - \Delta^3$ of order $\sim 1$ fm in the
proton rest frame. The calculation for valence d quarks gives much
worse results; the reasons for this discrepancy are discussed.  }

\vspace{1cm}
{\em Submitted to Phys.\ Rev.\ D}
\end{center}
\end{titlepage}

\newpage
%%%%%%GENERALITIES ABOUT DIS %%%%%%%%%%
\section{Introduction}
Deep inelastic lepton-hadron scattering has proved to be the best
testing ground for perturbative QCD. Thanks to the celebrated
factorization theorems \cite{collins1}, which can be derived quite
rigorously in this case by using the Operator Product Expansion (OPE),
the entire $Q^2$ dependence of the cross section can be calculated
perturbatively, while all dynamical effects of large distances are
included in a set of one-particle parton distribution functions given
at a certain reference scale.  Determination of the set of partonic
distributions --- quark, antiquark, and gluon --- is an ultimate goal
for the experimental studies of the deep inelastic scattering, and
also provides a challenging task for nonperturbative approaches to
QCD.

In the past ten years remarkable progress has been made at
experimental side, and apart from the region of small Bjorken x, there
is not much controversy regarding the existing parametrizations of
parton distributions. The theoretical progress has been much more
moderate. Apart from several quark-model or MIT bag model
calculations, there have been relatively few attempts to determine
parton distributions e.g., from QCD sum rules. The problem has
proved to be difficult for the theory. Purpose of this paper is to
point out that major part of theoretical problems in the calculations
of parton distributions is due to the fact that distribution functions
in momentum space for each particular value of the momentum
fraction x receive contributions from both small and large
longitudinal distances, which correspond to different physics and are
difficult to treat simultaneously. We argue that longitudinal {\em
distance} distributions are much easier from the theoretical point of
view than momentum distributions. They can be extracted from data
with marginal complications.  In what follows we shall demonstrate
that working with coordinate space distributions gives us a
selfconsistent formalism, which is not more complicated than the
standard one and its relation to the OPE is much closer. We illlustrate
the advantages of this approach by the calculation of valence quark
distributions in the framework of QCD sum rules.

Obviously, the longitudinal distance distributions are simply
Fourier transforms of the momentum distributions. For example for
the valence quark distribution one gets
\begin{equation}
Q_{\rm val}(z,\mu^2) = \int_0^1 du \cos{(uz)} q_V(u,\mu^2),
\label{FV}
\end{equation}
where $u$ is the momentum fraction.  The physical interpretation of
the variable z has been discussed in the literature since a long time
\cite{Iof0,Tera}.  In the center-of-mass (CM) system of the target a
deep-inelastic probe $\gamma^*$ (photon or neutrino) is converted into
a quark-antiquark pair at some space-time location. At large $Q^2$ the
pair travels with the speed of light along a light-like path,
interacts with the nucleon and is converted back into the probe. The
time interval between the conversion points $\gamma^* \to {\bar q} q$
and ${\bar q} q \to \gamma^*$ in the CM system, so called Ioffe time
$\tau_I$, measures the light-like distance essential for the
process. The Lorentz-invariant variable related to $\tau_I$ is denoted
by z. In the CM system the relation between these two variables takes
a simple form $\tau_I = \frac{z}{M}$, where M is the mass of the
target (nucleon).  For simplicity we shall call z just the Ioffe time.

\begin{figure}[h]
\centerline{
\epsfysize=0.7\textheight
\epsfbox{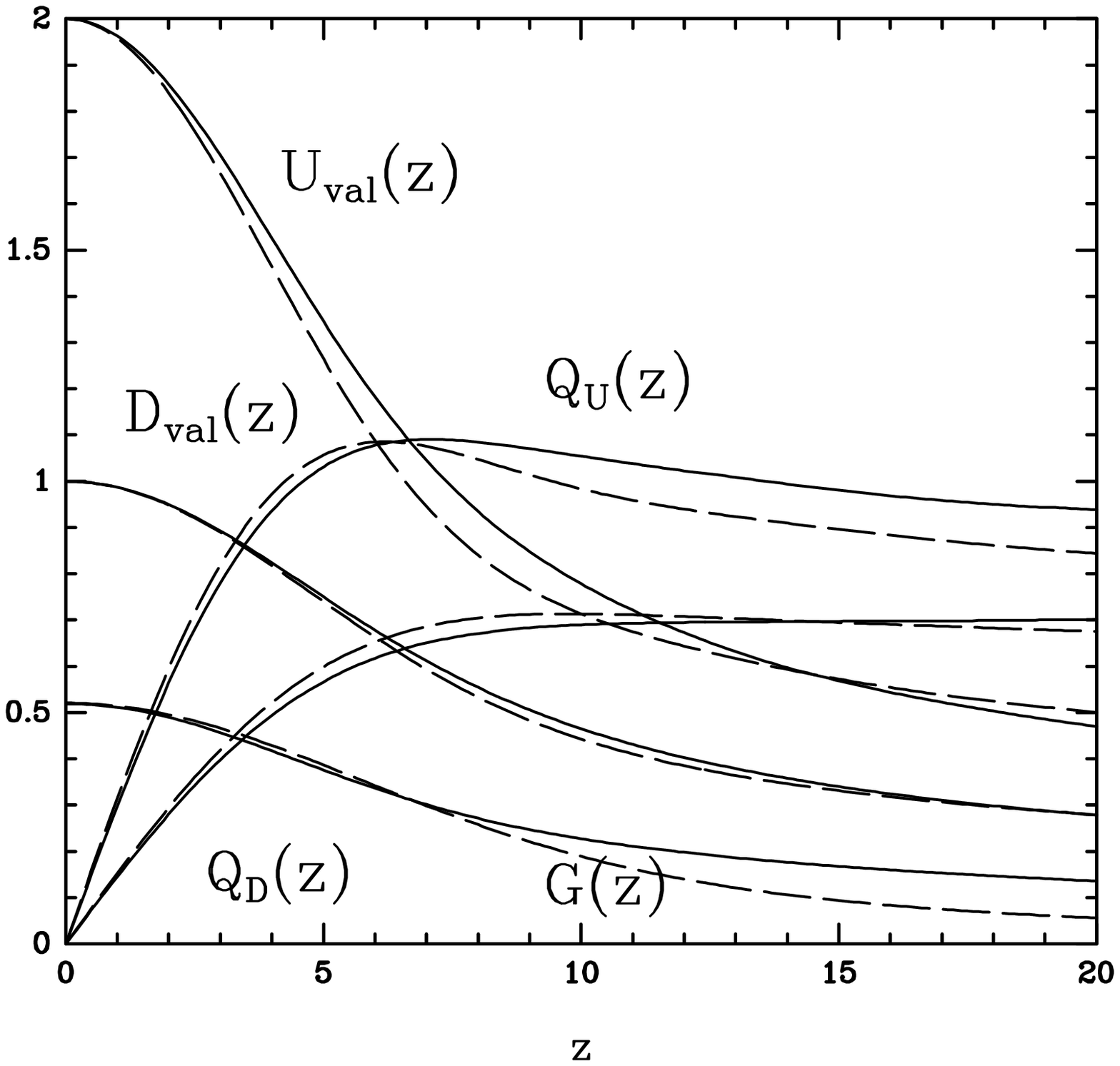}
}
\vspace{0.5in}
\caption[]{Ioffe time distributions for partons at the scale $\mu^2 = 4$
GeV$^2$. The solid line denotes Gl\"uck, Reya, Vogt (GRV)
\cite{GRVNLO} set of parametrizations, the dashed line denotes CTEQ
\cite{CTEQ} parametrization. $U_{\rm val}(z)$ and $D_{\rm val}(z)$ are
the valence quarks distributions, $Q_U(z)$ and $Q_D(z)$ are the C-even
up and down quark distributions, and $G(z)$ is the gluon distribution.
}\label{fig:1}

\end{figure}

\begin{figure}[h]
\centerline{
\epsfysize=0.7\textheight
\epsfbox{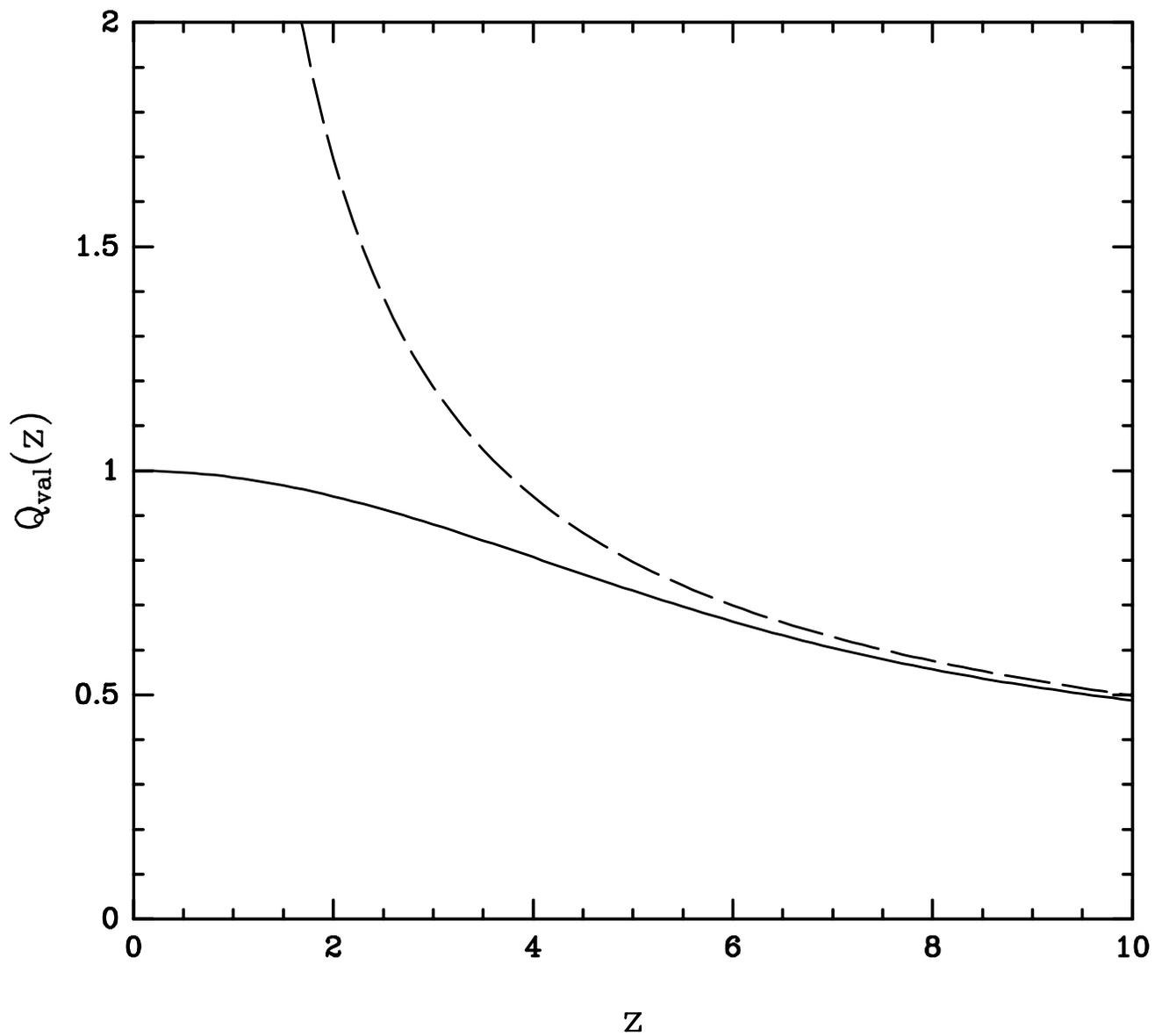}
}
\vspace{ 1.0in}
\caption[]{Model of the Ioffe time valence-type distribution corresponding
to the simple ansatz $q_V(u) = N u^{\alpha} (1-u)^\beta$ with $\alpha
= - 0.5$ and $\beta = 3$. The normalization is such that
$\int_0^1\, du q_V(u) = 1$. The dashed line shows the asymptotic
expansion (\protect{\ref{AsEx}}). Note that it matches almost
perfectly the true behaviour for $z \ge 6$.}\label{fig:2a}

\end{figure}

The idea to study z-distributions in order to understand the relative
importance of various light-cone distances for deep inelastic
scattering was proposed as early as in 1970 \cite{Tera}. There,  the
first analysis based on experimental data available at that time was
presented. The experimental information about the Ioffe-time
distributions available today is summarised in Fig.~\ref{fig:1}. We
plot Fourier transforms for two representative sets of
parametrizations \cite{CAT} -- the NLO Gl\"uck, Reya, Vogt
parametrization \cite{GRVNLO} and the CTEQ parametrization \cite{CTEQ}
-- for the valence quark ($u$ and $d$), quark plus antiquark, and
gluon distributions at $Q^2=4$ GeV$^2$. Exact definitions are given in
Sect.~2, where the connection of the Ioffe-time distributions and the
OPE is discussed in detail.  We note that all existing
parametrizations are in a fairly good agreement at small z, while at
large z there is some discrepancy. In order to get some insight about
the transition from the momentum to the coordinate space, we plot in
Fig.~\ref{fig:2a} the Ioffe-time valence quark distribution
corresponding to a simple ansatz
$$q_V(u) = N u^{\alpha} (1-u)^\beta$$
with $\alpha = - 0.5$ and $\beta = 3$, as suggested by the Regge theory
and the perturbative QCD arguments, respectively. The normalisation
factor $N$ ensures that $ Q_{\rm val}(0) = \int_0^1 du
q_V(u)~=~1$. The shape of $Q_{\rm val}$ depends very weakly on the
exact value of the parameter $\beta$ as long as $\alpha \ll \beta$.
The large-z behavior can readily be obtained from the standard
expressions for the asymptotic expansion of hypergeometric
functions.\footnote{ The asymptotical expansion contains in addition
oscillating terms $\sim 1/z^{\beta+1}$ but they can be neglected since
$\alpha \ll \beta$ in realistic case.}
\begin{equation}
 Q(z)   \sim   - \sin{({\pi\over 2}\alpha)} {\Gamma(\alpha+1)\over
z^{\alpha+1}} + \beta \cos{({\pi\over 2}\alpha)}
{\Gamma(\alpha+2)\over z^{\alpha+2}} + \cdots
\label{AsEx}
\end{equation}
Aa anticipated, the large-z asymptotics of
$Q(z)\sim z^{\alpha+1}$ is fixed by the Regge behaviour.
The valence quark distributions decrease at large
z as $\sim 1/\sqrt{z}$. The sea quark and gluon distributions
should approach a constant, or even may be rising functions of z at
large z.

An interesting question is at which values of z, or equivalently at
which longitudinal distances, the behavior of parton distributions is
already determined by the Regge asymptotics.  The dashed line on
Fig.~\ref{fig:2a} shows the asymptotic expansion of the function
$Q_{\rm val}(z)$ which matches almost perfectly the true behaviour
when $z \ge 6$. Because in the target (lab) frame $z = {1\over 2} M
\delta$, where $\delta$ denotes the distance along the light-cone, in
the case of the nucleon these values of z correspond to the light-cone
distances $\delta \ge 2.5$ fm.\footnote{To avoid confusion, note that
we are speaking here about {\em light-cone } distances, which in our
definition are factor two larger than physical longitudinal distances.}
  A nonperturbative calculation of the
parton distribution in this region is an extremely intricate
theoretical problem, which is essentially equivalent to providing a
nonperturbative input to the Lipatov's pomeron \cite{Lipatov} (or
Reggeon, in case of valence distributions). On the other hand,
calculation of the distributions at sufficiently small z may be within
reach of existing models, the QCD sum rules or the lattice
calculations.  Note that all parton distributions represented in
Fig.~\ref{fig:1} are very smooth at small z, which suggests that one
should be able to reproduce them in this region with only few terms in
the Taylor expansion around $z = 0$. These terms are related to the
first few moments of the momentum distributions, in other words to
nucleon expectation values of a few local operators of low dimension
(see below).  The most important question is whether there exists a
``matching window'', where both the Regge asymptotic formulae and the
small z expansion are applicable. Provided the answer is positive, one
could hope to get a quantitative description of the parton
distributions in the whole z range, matching these two different
inputs at a certain intermediate value of z - note
similarity with the usual QCD sum rule program.  Thus, the problem of
calculating the parton distributions can be posed as a problem of
calculating the distributions at distances of order $2-3$ fm at the
light-cone.  As we shall see below, the standard QCD sum rules are
sufficient for this purpose for the valence u-quark distributions, but
fail for the d-quarks.

\begin{figure}[h]
\centerline{
\epsfysize=0.7\textheight
\epsfbox{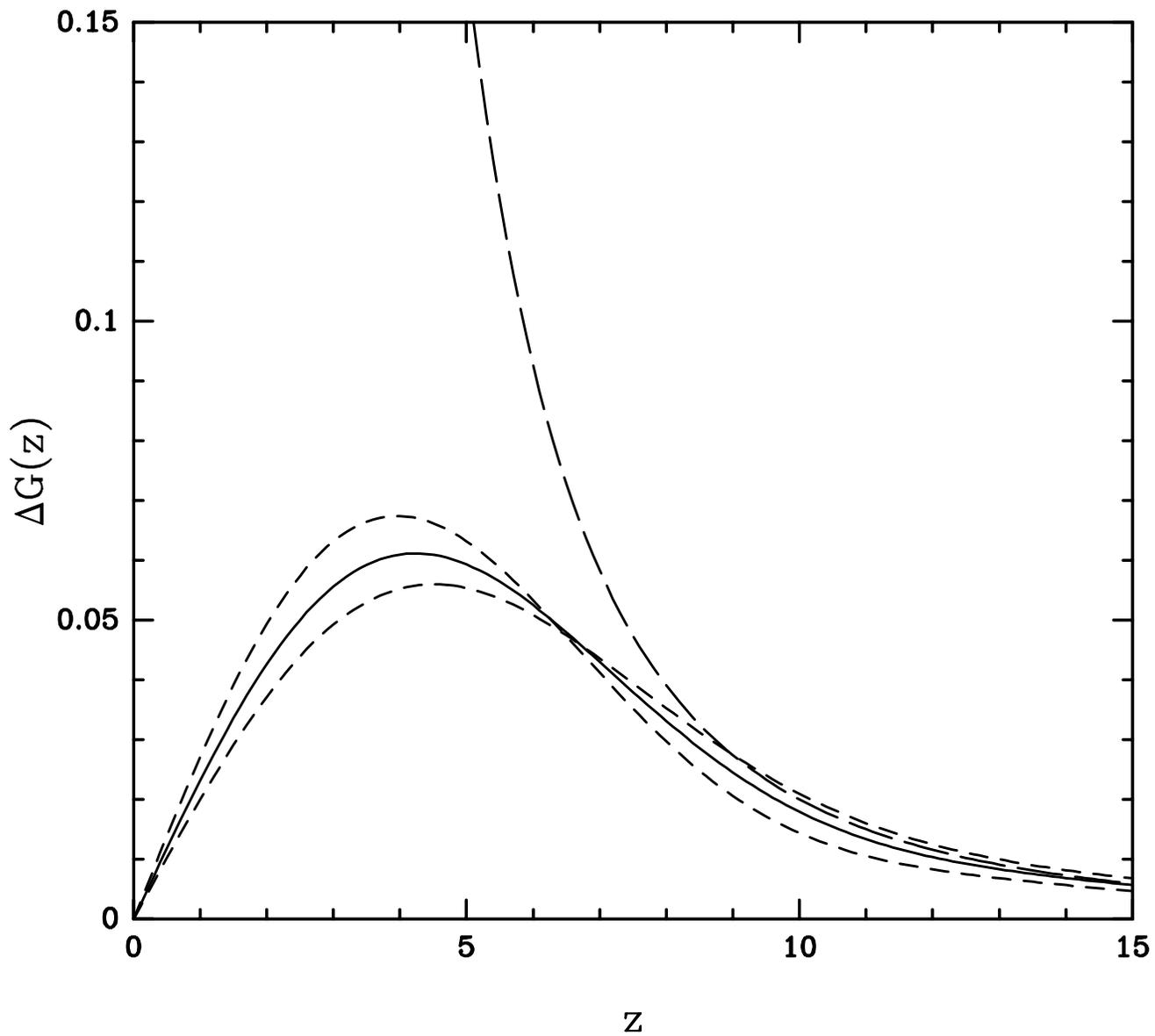}
}
\vspace{ 1.0in}
\caption[]{
Model of the Ioffe time distributions for polarized gluon density
$\Delta g(u) = N_G u^\alpha (1-u)^\beta$ with $\alpha = 0$ and $\beta
= 4$. The normalization constant $N_G$ is chosen in such a way that
the gluon polarisation $\Delta g = 0.5$. The two short-dashed curves
were obtained by taking $\beta$ equal to 3.5 and 4.5 respectively, and
keeping $\alpha=0$ and $\Delta g=0.5$ fixed. The long-dashed curve is
the asymptotic expansion (\protect{\ref{AsEx}}).
}\label{fig:2b}

\end{figure}

Playing around with typical parametrizations for parton distributions
which are used in modern experimental analysis, one can convince
oneself that in all cases the onset of the Regge behavior corresponds
to values of $z\sim 5-8$, see Fig.~\ref{fig:1}.  Another useful
example is given by the polarized gluon distribution. In this case we
define
\begin{equation}
\Delta G(z,\mu^2) = \int^1_0 du\,u \sin(uz) \Delta g(u,\mu^2)
\label{DG}
\end{equation}
where $\Delta g(u,\mu^2)$ is the usual polarized gluon distribution
depending on the momentum fraction $u$ (see Sect.~2 for details).
Note that the gluon polarisation
\begin{equation}
 \Delta g =\int_0^1 du\, \Delta g(u) =\int_0^\infty dz\,\Delta G(z)
\end{equation}
A typical shape of $ \Delta G(z)$ is shown in Fig.~\ref{fig:2b}.  It
has been obtained with a simple model of $\Delta g(u) = N_G u^\alpha
(1-u)^\beta$, and the solid curve corresponds to $\alpha = 0$ and
$\beta = 4$.  The normalization constant $N_G$ is chosen in such a way
that the gluon polarisation $\Delta g = 0.5$.  The two short-dashed
curves were obtained by taking $\beta$ equal to 3.5 and 4.5
respectively, and keeping $\alpha=0$ and $\Delta g=0.5$ fixed.  The
corresponding variation of $\Delta G(z)$ is rather mild, and one can
conclude that the behaviour of $\Delta g(u)$ at small u combined with
the value of $\Delta g$ determine to large extent  the shape of
$\Delta G(z)$. By the same argument, knowledge of $\Delta G(z)$ up to
the point of maximum, which is again at $z\sim 6$, is enough to
estimate the value of $\Delta g$ within, say, 50\% accuracy. Note
however that because of the more complicated shape of $\Delta G(z)$,
asymptotic expansion shown by the long-dashed curve in
Fig.~\ref{fig:2b} starts to be valid at larger values of $z \sim 10$.

Our presentation is organized as follows. In Sect.~2
we discuss the theoretical framework for the introduction of
parton distributions in coordinate space as matrix elements of
nonlocal operators, and we describe their  $Q^2$ evolution.
The presentation in this section mainly follows Refs.\cite{Collins,bal88}.
In Sec.~3 we give the QCD sum rule calculation of valence quark
distributions in coordinate space, and compare our approach to
the direct calculation in the momentum space in Ref. \cite{Iof1}.
Sect.~4 is reserved for a summary and conclusions. Some
technical details of the sum rule calculation are presented in the
Appendix.

%%%%%%%%%%%%%%%%%%%%%%%% SECTION 2 %%%%%%%%%%%%%%%%%%%%%%%%%%%%%%%%%%%%

\section{Covariant Definition of Ioffe Time Distributions
and their $Q^2$ Evolution}

An intuitive discussion of the space-time picture of deep
inelastic scattering in late \linebreak 60-s can be put on a
rigorous footing using the formalism of the Operator Product Expansion
(OPE). We are going to demonstrate that the Ioffe time distributions
arise naturally in this framework as reduced matrix elements of
nonlocal string operators on the light-cone.  Our presentation
essentially follows Refs. \cite{Collins,bal88}.

It is well known that the deep-inelastic $ep$ scattering
cross-section is related to the matrix element:
\begin{equation}
T_{\mu\nu}=
\int d^4 y\; \exp(iqy)
\langle P \mid T[j_\mu(y)^\dagger j_\nu(0)] \mid P \rangle
\label{Tmunu}
\end{equation}
where $\mid P\rangle$ represents a proton with momentum $P$ and $j$ is
an electromagnetic current operator.  This quantity describes hadronic
part of the process. The Operator Product Expansion applied to
$T[j(y)j(0)]$ gives rise to its systematic expansion in powers of the small
parameter ${\Lambda^2}/{Q^2}$ where $\Lambda$ is the QCD scale
of the order of 200 MeV and Q$^2$ = - q$^2$ is the virtuality of
the deep-inelastic probe. To the leading, twist-2 accuracy, i.e. when
all powers of $\Lambda^2/Q^2$ are neglected, the quark and
gluon operators appearing in the OPE of $T[j(y)j(0)]$ have the form:
\begin{eqnarray}
{\hat {\cal O}}_q^{\mu_1\dots\mu_n}(0) & = & \frac{1}{2} {\bar \Psi}(0)
\left\{ \gamma^{\mu_1} iD^{\mu_2}\dots iD^{\mu_n} \right\}_{\rm ST} \Psi(0)
\nonumber \\
{\hat {\cal O}}_g^{\mu_1\dots\mu_n}(0) & = & \frac{1}{2}\left\{
G^{\mu_1\nu}(0)iD^{\mu_2}\dots iD^{\mu_{n-1}} G_\nu^{\mu_n}(0)
\right\}_{\rm ST} \; ,
\label{QGop}
\end{eqnarray}
where $D^\mu = \partial^\mu - i g A^\mu$ denotes covariant derivative,
and $\Psi$ and $G^{\mu\nu}$ are the quark field and the gluon
field strength, respectively.  The subscript ST denotes the
symmetric and traceless part of a Lorentz tensor.

The operators defined in (\ref{QGop}) form an irreducible
representation of the Lorentz group. Their reduced matrix elements
$\langle \langle O_q^n \rangle \rangle$ and $\langle \langle O_g^n
\rangle \rangle$ are defined by the relations:
\begin{eqnarray}
\langle P \mid {\hat {\cal O}}_q^{\mu_1\dots\mu_n}(0) \mid P \rangle
& = &
\langle \langle O_q^n \rangle \rangle
\left\{ P^{\mu_1}\dots P^{\mu_n}\right\}_{ST} \nonumber \\
\langle P \mid {\hat {\cal O}}_g^{\mu_1\dots\mu_n}(0) \mid P \rangle
& = &
\langle \langle O_g^n \rangle \rangle
\left\{ P^{\mu_1}\dots P^{\mu_n}\right\}_{ST} \; .
\label{RME}
\end{eqnarray}
According to the standard analysis, the matrix elements are related to
the moments of familiar quark and gluon distribution functions $q(x)$,
${\bar q}(x)$ and $g(x)$.
\begin{eqnarray}
\langle \langle O_q^n \rangle \rangle_{\mu^2}  & = & \int_0^1 dx\, x^{n-1}
\left( q(x,\mu^2) + (-1)^n {\bar q}(x,\mu^2)\right)
\label{MOM1}\\
\langle \langle O_g^n \rangle \rangle_{\mu^2}  & = & \int_0^1 dx\,
x^{n-1} g(x,\mu^2) \; .
\label{MOM2}
\end{eqnarray}
Equation (\ref{MOM2}) holds only for even values of n. Note that
through renormalization the operators (\ref{QGop}) acquire a scale
dependence, which is related to the scale dependence of the parton
distribution in (\ref{MOM1}) and (\ref{MOM2}).

An alternative representation for relations (\ref{MOM2}) and
(\ref{MOM1}) has been noticed in \cite{Collins} long ago, and we want
to introduce it now. For that purpose let us define two light-like
vectors $n^\mu$ and $\Delta^\mu$, such that $n^2 = \Delta^2 = n \cdot
\Delta = 0$. Our convention is such that and for any vector $a$, $n
\cdot a \equiv a^+ = a^0 + a^3$. The vector $\Delta$ is just
proportional to $n$, $\Delta = \frac{1}{2}\delta n$, where $\delta
= \Delta^- = \Delta^0 - \Delta^3$ is the distance along the light cone.
Following Ref. \cite{Collins} we can write
\begin{eqnarray}
\langle P \mid {\bar \Psi}(\Delta) {\not\!  \,}
[\Delta;0] \Psi(0)
\mid P \rangle_{\mu^2} & = &  2 (P \cdot n) \int_0^1 du\,\Big[
(q(u,\mu^2)\exp{(iuz)} - {\bar q}(u,\mu^2)\exp{(-iu z)}\Big]
\nonumber \\
\label{REL1}
\end{eqnarray}
for quark distributions and
\begin{equation}
\langle P \mid G_{\mu\xi}(\Delta)[\Delta;0]G^\xi_\nu(0) \mid
P \rangle_{\mu^2} n^\mu n^\nu  =  4 (P \cdot n)^2 \int_0^1 du\,
g(u,\mu^2) u \cos{(u z)}
\label{REL2}
\end{equation}
for gluons.
In the above formulae $ z = P \cdot \Delta$ and we have
introduced the notation $[\Delta;0]$ for
the path-ordered exponential:
\begin{equation}
[\Delta;0] =
{\large P}\exp {\left[i g \Delta_\xi \int_0^1 ds A^\xi(\Delta s)\right]}
\label{POE}
\end{equation}
which is necessary for gauge independence of the parton distributions
considered. An easy way to obtain the relations (\ref{REL1}),
(\ref{REL2}) and (\ref{REL3}) is to insert the complete set of
intermediate light-cone quark or gluon states between the field
operators at the RHS working in the Schwinger gauge: $\Delta \cdot A
(\Delta) = 0$. Taylor expansion in $\Delta$ of both sides of
(\ref{REL1}) and (\ref{REL2}) gives exactly the set of relations
(\ref{MOM1}) and (\ref{MOM2}) between the matrix elements of local
operators and the moments of structure functions. Note that because
$n$ and $\Delta$ are light-like and proportional, the local
operators arising here automatically are of twist 2.

 Fourier transformation of (\ref{REL1}) and (\ref{REL2}) gives a gauge
invariant definition of parton momentum distributions in terms of
reduced matrix elements of leading twist nonlocal operators at the
light-cone \cite{Collins}. On the other hand, it is possible to
demonstrate that the usual program of the OPE can be formulated
directly in terms of nonlocal light-cone operators \cite{bal88}. Thus,
this formalism is consistent. For completeness, we quote the
definition for the polarised gluon distribution \cite{man90,bal91},
see discussion after Eq.(\ref{DG}):
\begin{equation}
\langle P, S \mid G_{\mu\xi}(\Delta)[\Delta;0]{\tilde G}^\xi_\nu(0)
\mid P,S \rangle_{\mu^2} n^\mu n^\nu  =  4i (P \cdot n) (S \cdot n)
\int_0^1 du\, \Delta g(u,\mu^2) u \sin{(u z)}\,,
\label{REL3}
\end{equation}
where $S$ is the nucleon spin vector normalized by $S^2 = - M_N^2$.

Taking C-odd and C-even combinations of the LHS of (\ref{REL1}) we
arrive at the definitions involving C-odd (valence) and C-even
combinations of parton densities:
\begin{equation}
\langle P \mid {\bar \Psi}(\Delta) {\not\! n \,} [\Delta;0] \Psi(0)
\mid P \rangle_{\mu^2} + (\Delta \to - \Delta)  =
4 (P \cdot n) \int_0^1 du\,
q_V(u,\mu^2) \cos{(z u)} \,
\label{REL1A}
\end{equation}
and
\begin{equation}
\langle P \mid {\bar \Psi}(\Delta) {\not\! n \,} [\Delta;0] \Psi(0)
\mid P \rangle_{\mu^2} - (\Delta \to - \Delta)  =
4 i (P \cdot n) \int_0^1 du
\left[q(u,\mu^2) + {\bar q}(u,\mu^2)\right]\sin{(z u)} \, , \nonumber \\
\label{REL1B}
\end{equation}
where $q_V(u,\mu^2) = q(u,\mu^2) - {\bar q}(u,\mu^2)$, and
${\not\! n \,} = n_\mu \gamma^\mu$.

As mentioned above, the conventional procedure is to Fourier
transform the above formulae ending up with the parton
distributions in momentum space. The main thrust of our paper is to
point out that the matrix elements appearing on the LHS of the
equations (\ref{REL1})--(\ref{REL1B}) have a clear physical
interpretation as the parton distributions in the longitudinal coordinates,
and are more adequate for the application of nonperturbative methods,
retaining at the same time the whole physical content of the momentum
space description. We define Ioffe time distributions by:
\begin{equation}
\langle P \mid {\bar \Psi}(\Delta) {\not\! n \,} [\Delta;0] \Psi(0)
\mid P \rangle_{\mu^2} + (\Delta \to - \Delta)  =
4 (P \cdot n)\,
Q_{\rm val}(z,\mu^2) \,,
\label{QQV}
\end{equation}
\begin{equation}
\langle P \mid {\bar \Psi}(\Delta)
{\not\! n \,} [\Delta;0] \Psi(0)
\mid P \rangle_{\mu^2} - (\Delta \to - \Delta)  =
4 i (P \cdot n)\, Q(z, \mu^2)
\label{QQQ}
\end{equation}
for quarks, and
\begin{equation}
\langle P \mid G_{\mu\xi}(\Delta)[\Delta;0]G^\xi_\nu(0) \mid
P \rangle_{\mu^2} n^\mu n^\nu  =  4 (P \cdot n)^2
G(z,\mu^2) \,,
\label{GGG}
\end{equation}
\begin{equation}
\langle P, S \mid G_{\mu\xi}(\Delta)[\Delta;0]{\tilde G}^\xi_\nu(0)
\mid P,S \rangle_{\mu^2} n^\mu n^\nu  =  4i (P \cdot n) (S \cdot n)
\Delta G(z,\mu^2) \,,
\label{Gdual}
\end{equation}
for nonpolarized and polarized gluon distributions, respectively.
Comparing to (\ref{REL1})--(\ref{REL1B}) we arrive at the relations
between the momentum and coordinate space distributions in (\ref{FV}),
(\ref{DG}).

According to the standard discussion the scale dependence of the
longitudinal momentum parton distributions is governed by GLAP equations
written in momentum space. However, it is also possible to
derive the corresponding evolution directly in coordinate
space \cite{Collins,bal88}. The coordinate-space version of the LLA
evolution equations has been obtained in
\cite{bal88} in the form of equations describing the
normalization-point dependence of the non-local operators (\ref{REL1})
and (\ref{REL2}). Taking the forward nucleon matrix element and
making use of relations (\ref{REL2}), (\ref{REL1A}) and (\ref{REL1B}) one
can derive evolution equations for Ioffe time distributions
$Q_{\rm val}(z,\mu^2)$, $Q(z,\mu^2)$ and $G(z,\mu^2)$.

To one loop accuracy the scale dependence of the valence quark Ioffe
time distribution (\ref{QQV}) is governed by
\begin{equation}
Q_{\rm val}(z;\mu_2^2) = Q_{\rm val}(z;\mu_1^2) - \frac{\alpha_S}{2 \pi} C_F
\ln{\frac{\mu_2^2}{\mu_1^2}} \int_0^1 du\, K(u) Q_{\rm val}(u z;\mu_1^2).
\label{EVNS}
\end{equation}
The kernel K(u) is given by
\begin{equation}
K(u)= \frac{1}{2} \delta({\bar u}) - {\bar u} - 2 \left[ \frac{u}{{\bar u}}
\right]_+
\label{KNS}
\end{equation}
where ${\bar u} = 1 - u$ and for any function f(u)
\begin{equation}
\int_0^1 du \left[ \frac{u}{{\bar u}} \right]_+ f(u) \equiv
\int_0^1 du\, \frac{u}{{\bar u}} (f(u) - f(1))
\label{plus}
\end{equation}

In the flavour-singlet channel the evolution mixes, as expected, quark
and gluon distributions $Q(z,\mu^2)$ (\ref{QQQ}) and $G(z,\mu^2)$
(\ref{GGG}):
\begin{equation}
\left[\!
\begin{array}{c}
Q(z;\mu_2^2) \\ z G(z;\mu_2^2)
\end{array}\! \right]
=
\left[\!
\begin{array}{c}
Q(z;\mu_1^2) \\ z G(z;\mu_1^2)
\end{array}\! \right]
- \frac{\alpha_S}{2 \pi} \ln{\frac{\mu_2^2}{\mu_1^2}} \int_0^1\! du
\left[\!
\begin{array}{cc}
C_F K_{QQ}(u) & N_f K_{QG}(u) \\
C_F K_{GQ}(u) & N_C K_{GG}(u)
\end{array}\! \right]
%\cdot
\left[\!
\begin{array}{c}
Q(uz;\mu_1^2) \\ z G(uz;\mu_1^2)
\end{array}\! \right]\, ,
\label{EVS}
\end{equation}
where
\begin{eqnarray}
K_{QQ}(u) & = & \frac{1}{2} \delta({\bar u}) - {\bar u} -
2 \left[ \frac{u}{{\bar u}}
\right]_+
\nonumber \\
K_{GG}(u) & = &
\left(\frac{1}{6} + \frac{1}{3}\frac{N_f}{N_C}\right) \delta({\bar u})
- 2 \left[ \frac{u}{{\bar u}} \right]_+ + 2 (u^3 - {\bar u}^2)
\nonumber \\
K_{QG}(u) & = & - \frac{1}{3} {\bar u}(2 {\bar u}^2 + 3 u)
\nonumber \\
K_{GQ}(u) & = & - \delta({\bar u}) - 2 {\bar u} \; ,
\label{KS}
\end{eqnarray}
where $N_C$ is the number of colours and $N_f$ is the number of active
flavours. Equations (\ref{EVNS}) and (\ref{EVS}) allow for a
systematic study of the evolution of Ioffe time parton densities
exactly in the same manner as the conventional GLAP equations do for
the longitudinal momentum parton densities. Indeed, let us rewrite
(\ref{EVNS}) and (\ref{EVS}) as the evolution equations for flavour
non-singlet and flavour singlet distributions:
\begin{equation}
\mu^2 \frac{\partial}{\partial \mu^2}\, Q_{\rm val}(z;\mu^2) =
- \frac{\alpha_S(\mu^2)}{2 \pi} C_F
\int_0^1 du\, K(u) Q_{\rm val}(u z;\mu^2).
\label{EVNS1}
\end{equation}
and
\begin{equation}
\mu^2 \frac{\partial}{\partial \mu^2}\,
\left[
\begin{array}{c}
Q(z;\mu^2) \\ z G(z;\mu^2)
\end{array} \right]
=
- \frac{\alpha_S(\mu^2)}{2 \pi}  \int_0^1 du
\left[
\begin{array}{cc}
C_F K_{QQ}(u) & N_f K_{QG}(u) \\
C_F K_{GQ}(u) & N_C K_{GG}(u)
\end{array} \right]
%\cdot
\left[
\begin{array}{c}
Q(uz;\mu^2) \\ z G(uz;\mu^2)
\end{array} \right]\; .
\label{EVS1}
\end{equation}
which can be identified as the RGE for twist-2 Ioffe time distributions.

A beautiful feature of the RGE equations (\ref{EVNS1}) and
(\ref{EVS1}) is that they explicitely show the relevance of the short
distance expansion: To calculate the QCD evolution of the
distributions, one needs to know them at a certain reference scale at
{\em smaller } values of the Ioffe time. Stated differently, the
evolution equation for the nonlocal operators
(\ref{REL1})--(\ref{REL3}) involves these operators at quark-antiquark
distances smaller than the initial separation.  This is in contrast to
the evolution of fragmentation functions, which is essentially
nonlocal in the coordinate space \cite{BB91}.

As it is well known, the integrodifferential equations (\ref{EVNS1})
and (\ref{EVS1}) can be transformed into ordinary differential
equations by Mellin transformation
$$F(z) \to {\hat F}(\nu) = \int_0^\infty dz\, z^{\nu-1} F(z) \; .$$
As a consequence we obtain in the flavour non-singlet case:
\begin{equation}
\mu^2 \frac{\partial}{\partial \mu^2}\, {\hat Q}_{\rm val}(\nu;\mu^2)
= - \frac{\alpha_S(\mu^2)}{4 \pi} \gamma(-\nu - 1)
{\hat Q}_{\rm val}(\nu;\mu^2)
\label{RGENS}
\end{equation}
where we have introduced the function
\begin{equation}
\gamma(\nu) = 2 C_F \int_0^1 du\, K(u) u^{\nu-1}
\end{equation}
Equation (\ref{RGENS}) has the well known solution
\begin{equation}
{\hat Q}_{\rm val}(\nu;\mu_2^2) = \left(
\frac{\alpha_S(\mu_2^2)}{\alpha_S(\mu_1^2)}\right )^{\gamma(-\nu+1))/b}
{\hat Q}_{\rm val}(\nu;\mu_1^2)\; ,
\label{RGENS1}
\end{equation}
where in QCD $b = \frac{11}{2} - \frac{1}{3} N_f$.
The Ioffe time distribution at the scale
$\mu_2^2$ can be obtained with the help of the inverse Mellin transformation:
$${\hat F}(\nu) \to F(z) = \frac{1}{2\pi i}\int_{c-i \infty}^{c+i
\infty} d\nu\, z^{-\nu} F(\nu) \, .$$
Explicitely, one gets \cite{bal88}
\begin{equation}
{\hat Q}_{\rm val}(z;\mu_2^2) = \int_{-\infty}^\infty \frac{d\nu}{2\pi}
\left(
\frac{\alpha_S(\mu_2^2)}{\alpha_S(\mu_1^2)}\right )^{\gamma(1/2-i\nu)/b}
\int_0^\infty du\, u^{i\nu - 1/2} {\hat Q}_{\rm val}(uz;\mu_1^2) \, .
\label{RGENS2}
\end{equation}

\begin{figure}[h]
\centerline{
\epsfysize=0.7\textheight
\epsfbox{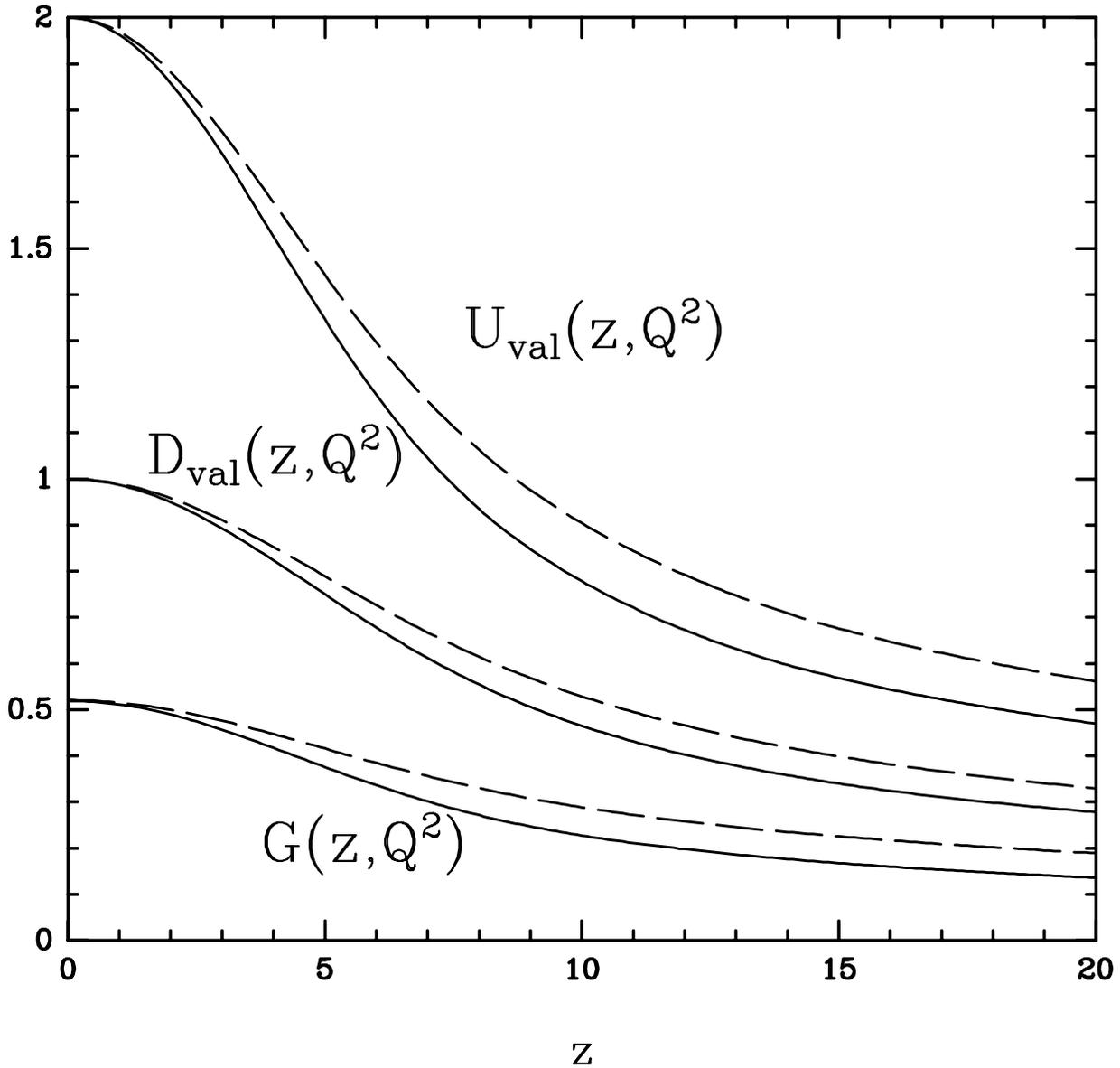}
}
\vspace{ 1.0in}
\caption[]{
Evolution of valence $U_{\rm val}(z,Q^2)$ and $D_{\rm val}(z,Q^2)$,
and gluon $G(z,Q^2)$ Ioffe time distributions. GRV \cite{GRVNLO}
parametrisation has been used. The solid line corresponds to $Q^2 = 4$
GeV$^2$, the dashed line to $Q^2 = 20$ GeV$^2$, respectively.
}\label{fig:3a}

\end{figure}

\begin{figure}[h]
\centerline{
\epsfysize=0.7\textheight
\epsfbox{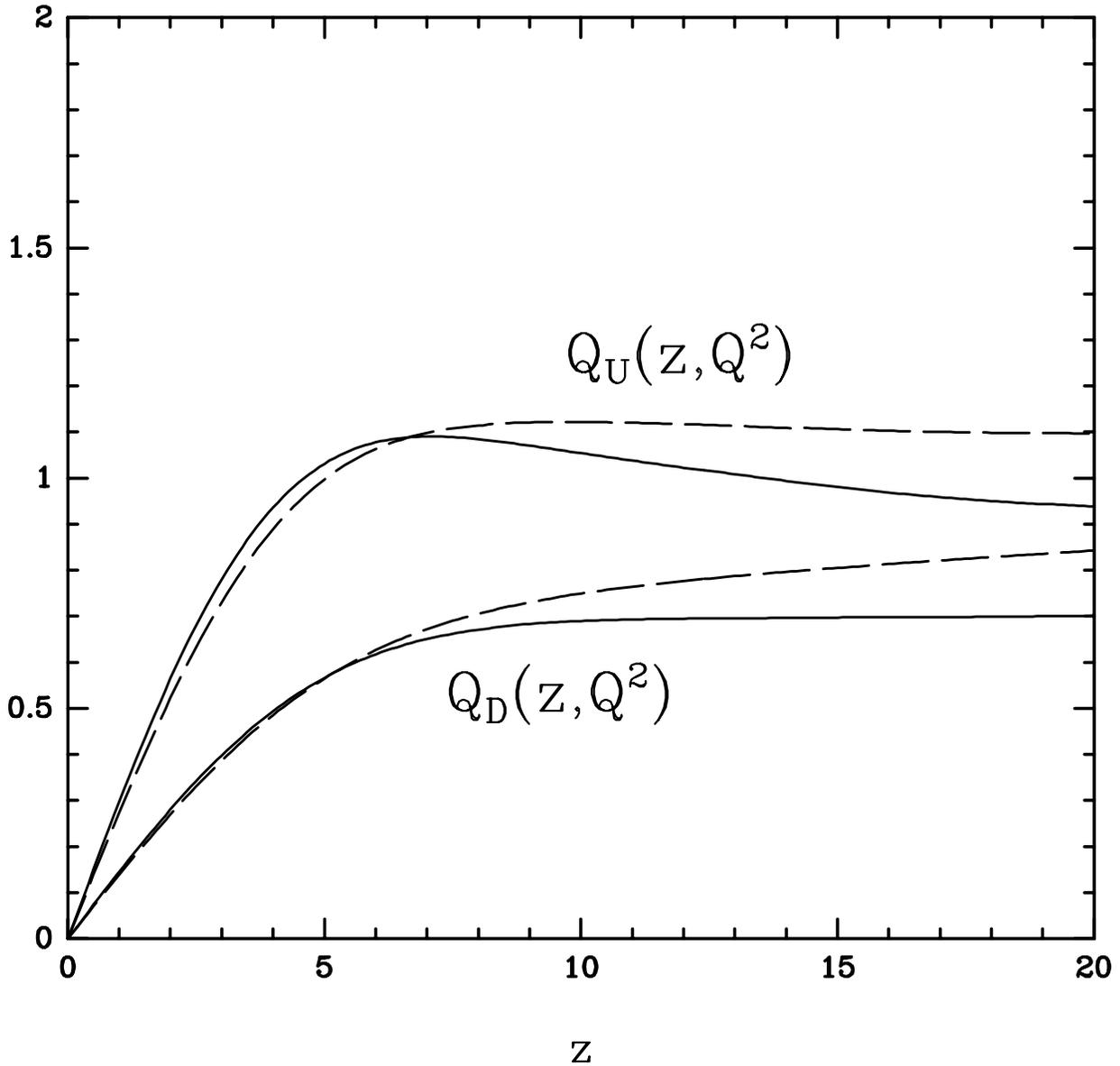}
}
\vspace{ 1.0in}
\caption[]{
Evolution of up and down quark C-even $Q_U(z,Q^2)$ and $Q_D(z,Q^2)$
Ioffe time distributions. GRV \cite{GRVNLO} parametrisation has been
used. The solid line corresponds to $Q^2 = 4$ GeV$^2$, the dashed line
to $Q^2 = 20$ GeV$^2$, respectively.
}\label{fig:3b}

\end{figure}

Typical results of the low-scale evolution i.e., in the low $\mu^2$
range are illustrated on Figures \ref{fig:3a} and \ref{fig:3b}. Figure
\ref{fig:3a} shows valence and gluon GRV \cite{GRVNLO} Ioffe time
distributions evolved between $\mu^2$ = 4 and 20 GeV$^2$. Figure
\ref{fig:3b} shows the same for u- and d-quark distributions. Note
that in this range of scales the evolution affects mainly the large-z
behavior of the distributions.

%%%%%%%%%%%%%%%%%%%%%%% SECTION 3  %%%%%%%%%%%%%%%%%%%%%%%%%%%%%%%%%

\section{ Valence Quark Distributions from QCD Sum Rules}

We can now summarize our discussion in the following way. We have
analysed the reduced matrix elements of QCD string operators of twist
2 as a function of the light-like separation between fields. We have
found very smooth behaviour which makes such objects convenient for
theoretical studies. It can be also shown that the large separations
are dominated by the asymptotics of the corresponding structure
function at small values of Bjorken x. Once this asymptotics is
known from e.g. the Regge arguments, the remaining non-trivial
information is contained in the domain of moderately large
separations. This region could be accesible to presently developed
analytical  methods like QCD sum rules, instanton models
of QCD vacuum \cite{shuryak}, or lattice calculations.

In this paper a QCD sum rule calculation is carried out for the
valence quark distributions.  In the last decade the QCD sum rule approach
has been applied succesfully to a variety of problems, including
estimation of hadron masses and couplings, elastic and transition form
factors etc. In the context of this paper it is necessary to mention the
calculation of the fraction of proton momentum carried by gluons in
Ref. \cite{kolesnichenko,BelBlok} and the calculation of structure
functions at intermediate values of Bjorken variable in
Ref. \cite{Iof1}. In what follows, we shall often refer to this latter
analysis to compare the calculations in coordinate and momentum space.

The basic idea of the QCD sum rule technique is to use duality
between hadronic and partonic representations of a suitable correlation
function to extract the quantity of interest by requiring that the
two descriptions match each other at intermediate scales.
To calculate the valence quark distributions, we choose to work with
\begin{equation}
\Pi^{u,d} =
i^2 \int d^4x\, d^4y \exp{(i p\cdot x + i q \cdot y)}
\langle \Omega \mid T [\eta(x) {\bar \eta}(0)
{\hat O}^{u,d}_S (y+\frac{\Delta}{2};y-\frac{\Delta}{2})]
 \mid \Omega \rangle
\label{PiQ}
\end{equation}
where
\begin{equation}
{\hat O}^i_S(y+\frac{\Delta}{2};y-\frac{\Delta}{2})
 \equiv {\bar \Psi}^i(y+\frac{\Delta}{2})\gamma_\mu n^\mu
[y+\frac{\Delta}{2};y-\frac{\Delta}{2}] \Psi^i(y-\frac{\Delta}{2}) +
(\Delta \to -\Delta)\,.
\label{OurOp}
\end{equation}
Here $i=u,d$ denotes quark flavor, $n^\mu=(1,0,0,-1)$ is the "unit"
light-like vector,$n^2 = 0$, and the splitting $\Delta^\mu$ is
light-like and proportional to $n^\mu$.

Finally, $\eta(x)$ is the standard
 interpolating current for the proton \cite{Iof2}
\begin{equation}
\eta(x) =
\epsilon^{abc} u^a(x)^T C \gamma_\mu u^b(x) \gamma_5 \gamma^\mu d^c(x)
\label{Eta}
\end{equation}
where u(x) and d(x) denote u- and d-quark fields, respectively and
$a,b,c$ are color indices.

According to Eq.~(\ref{QQV}) the proton matrix element of ${\hat
O}^i(y+\frac{\Delta}{2};y-\frac{\Delta}{2})$ defines the valence quark
distribution. In the following we choose a special kinematics taking the
momentum transfer $q_\mu$ to be light-like $q^2=0$ and orthogonal to
the interquark separation i.e., $\Delta \cdot q = 0$.  In this
case, the nucleon contribution to the correlation function (\ref{PiQ})
can be extracted in the form
\begin{equation}
\frac{1}{4} {\rm Tr}  {\not\! n \,}
\Pi^{u,d} = \frac{4 \; \lambda_N^2}
{(p_1^2-M_N^2)(p_2^2-M_N^2)} (p\cdot n)^2 \int_0^1 du\,
q_V^{u,d}(u)\cos{(u p\cdot\Delta)} + {\rm continuum}
\label{PiQ1}
\end{equation}
where  $p_1^2 = p^2$ and $p_2^2 =(p+q)^2$.
The coupling
 $\lambda_N$ is defined by
$$
\langle \Omega \mid \eta(0) \mid P, N \rangle = \lambda_N u_N(P)
$$
where $u_N(P)$ is the nucleon spinor.

Note that the operator $\hat O$ in (\ref{OurOp}) is essentially the
point-splitted vector current. In the limit $\Delta=0$ a Ward identity
relates the three-point correlation function (\ref{PiQ}) to the
derivative of the two-point correlation function of two nucleon
currents
\begin{eqnarray}
 \Pi^{u,d}(\Delta=0) & = & N^{u,v}
 \int_0^1 dv\, n_\mu
\frac{\partial}{\partial { p}_\mu}
\Pi^{(2)}({ p+vq})\, ,
\nonumber\\
\Pi^{(2)}(p) &=&
i \int d^4 x \exp{(i { p}\cdot x)}
\langle \Omega \mid T\left[ \eta(x)
{\bar \eta}(0) \right] \mid \Omega \rangle\,,
\label{W5}
\end{eqnarray}
where $N^u=2$, $N^d=1$ are the numbers of valence quarks in the
proton. The derivation of (\ref{W5}) and of a more general Ward
identity for arbitrary separation $\Delta$ is given in the Appendix.
Note that
\begin{equation}
\frac{1}{4} {\rm Tr} {\not\! n \,} \Pi^{(2)}(p) =
\frac{\lambda_N}{M_N^2 - p^2}(p \cdot n)  + {\rm continuum}.
\label{MN}
\end{equation}
Substituting (\ref{MN}) in (\ref{W5}) and comparing to (\ref{PiQ1})
one obtains the normalization conditions
\begin{eqnarray}
U_{\rm val}(0) &=& \int_0^1 du\, q_u(u) = N^u = 2\,,
\nonumber\\
D_{\rm val}(0) &=& \int_0^1 du\, q_d(u) = N^d = 1\,,
\label{NQ}
\end{eqnarray}
which are exact in the QCD sum rule approach, provided the Ward
identity is not spoiled by the continuum subtraction (see below).

Main task is the calculation of the correlation function (\ref{PiQ})
in QCD. If  both $p^2$ and $q^2$ are sufficiently large (and negative)
the dominant contributions come from small distances
 $x$ and $y$ of order
${1}/{-p^2}$ and ${1}/{-q^2}$, respectively.
Thus the standard machinery of the short-distance expansion
is applicable, allowing to express the result as a power series
in terms of vacuum quark and gluon condensates. In the case of
 the forward matrix elements (i.e., for $q^2 =0$)
 the situation is more involved because the relevant
distances in the t-channel are not constrained by the external momenta
 and can be arbitrarily large.

The solution to this problem was first formulated by Balitsky
\cite{Bal1}. The Operator Product Expansion of the correlation
functions of the type (\ref{PiQ}) has a twofold structure. Terms of
the first type come from the region x$^2 \sim$ y$^2 \sim {1}/{-p^2}$
and are proportional to vacuum expectation values (VEV) of local gauge
invariant operators, multiplied by coefficient functions depending on
$p_1^2$, $p_2^2$.  In the following we refer to these terms as to
local power corrections (LPC).

Terms of the second type are called bilocal power corrections (BPC)
and correspond to the contributions of large $y^2\gg x^2\sim 1/p^2$.
In order to treat these terms for arbitrary $q$ in the ``Bjorken
limit'' $p^2\sim q\cdot p \rightarrow \infty$ one should expand the
T-product of nucleon currents
\begin{equation}
T[\eta(x)\eta(0)] = \sum_n C_n^{BL}(x^2) {\hat S}^n(x;0) ,
\label{BPC}
\end{equation}
in a  series of non-local,
gauge-invariant ``string'' operators of increasing
twist n ${\hat S}^n(x;0)$, cf. \cite{bal88}. This expansion
can  be inserted into the correlation function (\ref{PiQ}),
producing a power series in $1/p^2$ with coefficients given by
the correlation functions of two nonlocal light-cone operators
\begin{equation}
i \int dy \exp{(i q\cdot y)} \langle \Omega \mid
T\left[{\hat O}^{u,d}(y+\frac{\Delta}{2};y-\frac{\Delta}{2})\,
 {\hat S}^n(x;0)\right] \mid \Omega \rangle .
\label{BPC1}
\end{equation}
In general, calculation of the correlator (\ref{BPC1}) requires
construction of a specific sum rule and may be very complicated.
Remarkably, we have found that the most important BPC of dimension 6
can be evaluated exactly (i.e. related to the quark condensate) by
using the equations of motion. The derivation essentially uses the
Ward identity which we obtain in the Appendix.

It should be noted that the OPE for the correlation function
(\ref{PiQ}) for $q^2 =0$ is given by the sum of both LPC's and
BPC's. In general only this sum has a physical meaning and is
regularisation scheme independent.

Apart from these general remarks, we shall not go into details of the
calculation which is relatively straightforward.
A few more remarks are necessary, however, concerning the specific
techniques used in the QCD sum rules approach to suppress contributions of
higher states and further taking them into account
in the duality approximation.

To this end, the Borel transformation is applied to both sides of the
sum rule, improving the convergence of the operator product expansion
series and suppressing exponentially contributions of higher resonances.
The rationale for keeping nonzero value of the momentum transfer $q$
in the above discussion is that in this kinematics one can consider
$p_1^2=p^2$ and $p_2^2 = (p+q)^2$ as independent variables and perform
the Borel transformation in both momenta. The advantage of this
procedure is that in the double dispersion relation
\begin{equation}
\Pi = \int_0^\infty \int_0^\infty ds_1 ds_2
\frac{\rho(s_1,s_2)}{(s_1-p_1^2)(s_2-p_2^2)} + \dots ,
\label{Mand}
\end{equation}
where $\rho(s_1,s_2)$ denotes the spectral density, one can ignore
subtraction terms and contributions corresponding to non-diagonal
transitions with singularities in only one of the two variables. In
the standard duality approximation higher resonances and the continuum
contribution are taken into account by the following
model for the {\em physical} spectral density:
\begin{equation}
\rho(s_1,s_2) = \lambda_N^2 \delta(s_1 - M_N^2) \delta(s_2 - M_N^2)
\langle N, p_1 \mid {\hat O}^{u,d}
 \mid p_2, N \rangle +
\Theta(s_1 - s_0) \Theta(s_2 - s_0) \rho_c(s_1,s_2) .
\label{rho}
\end{equation}
where $ \rho_c(s_1,s_2)$ is the  corresponding spectral density
 {\em calculated} in perturbative QCD.
Thus, by assumption, subtraction of the continuum contribution
corresponds to constraining the integration region in $(s_1,s_2)$-plane
to the duality region $s_1,s_2 <s_0$.

In the theoretical part of the sum rule the double Borel transformation
is performed using the following general formula
\begin{equation}
 B \left\{ \frac{\Gamma(\nu)}
{[- {\bar v} p_1^2 - v p_2^2]^\nu} \right\} =
t^{2-\nu} \delta\left(v - \frac{M_1^2}{M_1^2+M_2^2}\right)
\label{Borel0}
\end{equation}
where $t$ denotes the symmetric
combination of Borel parameters $M_1^2$ and $M_2^2$
\begin{equation}
 t  \equiv \frac{M_1^2 M_2^2}{M_1^2+M_2^2}
\label{Borel}
\end{equation}
with ${\bar v} = 1- v$. In the symmetric case $M_1^2=M_2^2 = 2 t$
the subtraction of the continuum contribution corresponds to the
replacement
\begin{eqnarray}
t^n &\to & t^n E_n(t,s_0)\,;
\nonumber\\
  E_n(t,s_0)& =&
\left (1-e^{-s_0/t}\left[ 1+ \frac{s_0}{t}+\ldots +
\frac{1}{(n-1)!}
\left(\frac{s_0}{t}\right)^{n-1}\right]\right).
\label{AEn}
\end{eqnarray}
in all terms of the OPE containing positive powers of the Borel
parameter $t$.

\begin{figure}[h]
\centerline{
\epsfysize=0.9\textheight
\epsfbox{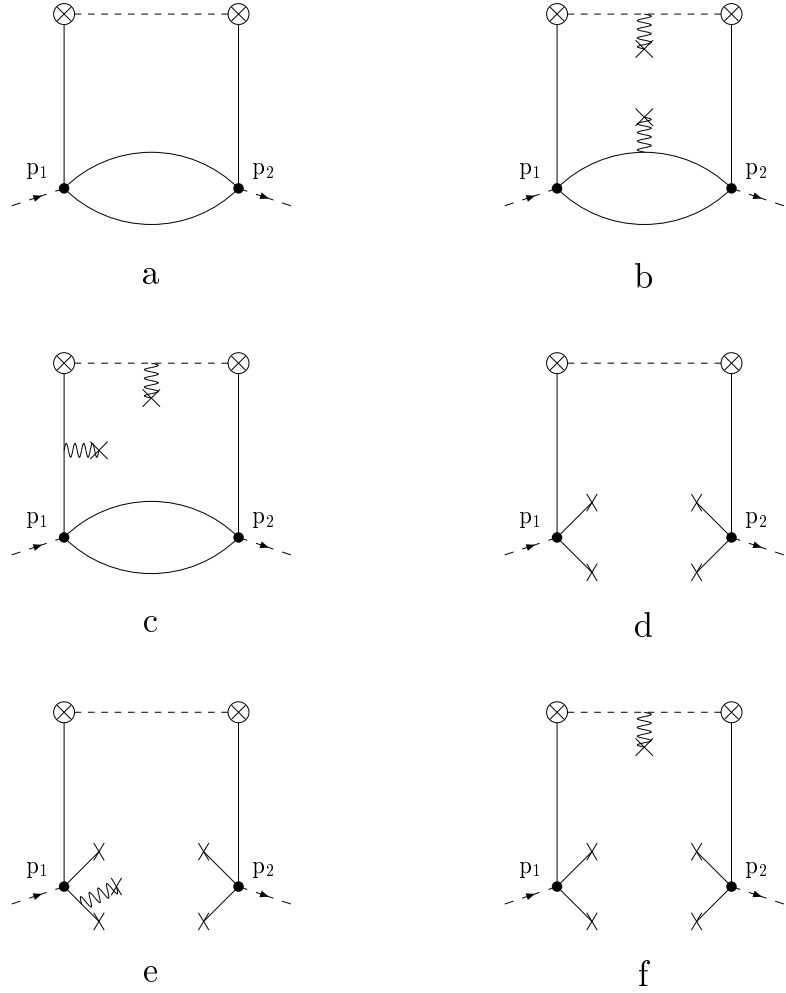}
}
\vspace{ - 0.8in}
\caption[]{
Typical diagrams contributing to the OPE of the correlator
(\protect{\ref{PiQ}}).  }\label{fig:4}

\end{figure}

\begin{figure}[h]
\centerline{
\epsfysize=1.5\textheight
\epsfbox{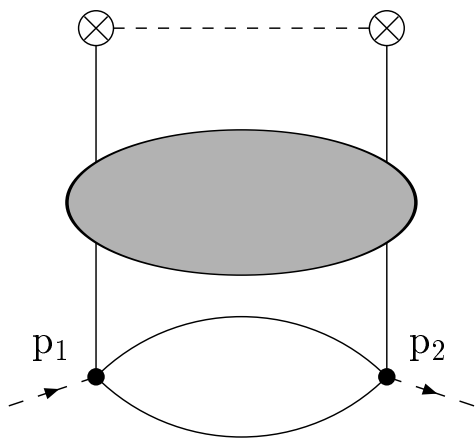}
}
\vspace{ - 7.5in}
\caption[]{
Generic form of a bilocal correction in the OPE of the
correlator (\protect{\ref{PiQ}}).
}\label{fig:5}

\end{figure}

An explicit calculation leads to the following sum rule for the
valence quark distributions:
\begin{eqnarray}
e^{-M_N^2/t} \lambda_N^2
\left\{\begin{array}{c}U_{\rm val}(z)\\ D_{\rm val}(z)\end{array}\right\}
 &= &
 f_{pert}^{u,d}(z,t,s_0) +
 f_{4}^{u,d}(z,t,s_0)\langle (\alpha_s/\pi)G^2\rangle +
f_{6}^{u,d}(z,t)\langle \bar q q\rangle^2 \nonumber\\
&&{}+
f_{8}^{u,d}(z,t)\langle \bar q q\rangle\cdot
\langle\bar q \sigma g G q\rangle +\ldots
\label{AOPE}
\end{eqnarray}
where we keep contributions of perturbation theory and operators up to
dimension 8.  The coefficient functions are given by diagrams shown in
Fig.~\ref{fig:4}.  For the bilocal power corrections, examplified in
Fig.~\ref{fig:5}, we take into account all contributions related to
contact terms, and neglect contributions of correlation functions
involving explicitly the gluonic fields in addition to quark fields
which arise from the first term in the Ward identity in (\ref{W3}).

We find:
\begin{eqnarray}
f_0^u(z) &=& \frac{1}{16 \pi^4} t^3 E_3(t,s_0) \int_0^1 du (9 u {\bar u}^2 +
{\bar u}^3) \cos{(u z)} \nonumber \\
f_0^d(z) &=& \frac{1}{16 \pi^4} t^3 E_3(t,s_0) \int_0^1 du (3 u {\bar u}^2 +
{\bar u}^3) \cos{(u z)} \\
\label{PT}
f_4^u(z) &=& \frac{1}{96 \pi^3} t E_1(t,s_0) \int_0^1 du [(4 \delta(u) +
6 u - 1 + \frac{1}{3} z^2 {\bar u}^3) \cos{(u z)} + z {\bar
u}^3 u^{-1} \sin{(u z)}] \nonumber \\
f_4^d(z) &=& \frac{1}{96 \pi^3} t E_1(t,s_0) \int_0^1 du [(2 \delta(u) +
3 {\bar u} - u + \frac{1}{3} z^2 {\bar u}^3) \cos{(u z)} + z {\bar
u}^3 u^{-1} \sin{(u z)}] \nonumber \\
\\
\label{DIM4}
f_6^u(z) &=& \frac{4}{3} \nonumber \\
f_6^d(z) &=& \frac{2}{3} \cos{z} \\
\label{DIM6}
f_8^u(z) &=& -\frac{4}{9} t^{-1} \nonumber \\
f_8^d(z) &=& t^{-1} ( - \frac{2}{9} \cos{z} + \frac{19}{54} z \sin{z})
\\ \nonumber
\label{DIM8}
\end{eqnarray}
To keep the correct normalization of parton densities (\ref{NQ}) the
coupling $\lambda_N^2$ in (\ref{AOPE}) should be substituted by the
corresponding sum rule \cite{Iof2}
\begin{equation}
e^{-M_N^2/t} \lambda_N^2  =
\frac{1}{32 \pi^4} t^3 E_3(t,s_0) +
\frac{1}{32 \pi^3} t E_1(t,s_0) \langle (\alpha_s/\pi)G^2\rangle +
\frac{2}{3} \langle \bar q q\rangle^2 -
\frac{2}{9} t^{-1} \langle \bar q q\rangle\cdot
\langle\bar q \sigma g G q\rangle +\ldots
\label{two-point}
\end{equation}
keeping the same terms in the OPE and using the same values of the
Borel parameter $t$ and the continuum threshold $s_0$ as in
(\ref{AOPE}). Note that the coefficients in front of the dimension 8
terms in (49) and (\ref{two-point}) differ slightly from the
corresponding ones in \cite{Iof1} and \cite{Iof2}. The reason is that
the authors of \cite{Iof1,Iof2} evaluate the vacuum expectation values
of nonlocal operators such as ${\bar \Psi}(0)\Psi(x) {\bar \Psi}(0)
\Psi(x)$ reducing them to $\sim (\langle {\bar \Psi}(0)\Psi(x)\rangle)^2$
using the hypothesis of the dominance of the vacuum
intermediate state, while we use this hypothesis in the calculation of
local operators, like ${\bar \Psi}D^2\Psi {\bar \Psi} \Psi$ only. This
leads to extra terms as compared to \cite{Iof1,Iof2}. These terms are,
however, suppressed as $\frac{1}{N_C}$, N$_C$ being the number of colors,
and the difference is not important numerically.

Please note that all coefficient functions in the sum rule in
(\ref{AOPE}) are smooth functions of the Ioffe time, in contrast to
the sum rules for the momentum fraction distributions given in
Ref. \cite{Iof1} which contain expansions involving singular
functions. Indeed, making a Fourier transform of our expressions, we
obtain the following sum rule for momentum fraction distributions of
valence quarks:
\begin{eqnarray}
e^{-M_N^2/t} \lambda_N^2 q^{u,d}_V(u) &= &
 \tilde f_{pert}^{u,d}(u,t,s_0) +
 \tilde f_{4}^{u,d}(u,t,s_0)\langle (\alpha_s/\pi)G^2\rangle +
\tilde f_{6}^{u,d}(u,t)\langle \bar q q\rangle^2 \nonumber\\
&&{}+
\tilde f_{8}^{u,d}(u,t)\langle \bar q q\rangle\cdot
\langle\bar q \sigma g G q\rangle +\ldots
\end{eqnarray}
with the coefficients:
\begin{eqnarray}
\tilde f_0^u(u) &=&
\frac{1}{16 \pi^4} t^3 E_3(t,s_0) (9 u {\bar u}^2 +
{\bar u}^3) \nonumber \\
\tilde f_0^d(u) &=&
\frac{1}{16 \pi^4} t^3 E_3(t,s_0) (3 u {\bar u}^2 +
{\bar u}^3)  \\
\label{PTu}
\tilde f_4^u(u) &=&
\frac{1}{96 \pi^3} t E_1(t,s_0) [(3 \delta(u) +
6 u - \left[ \frac{1}{u^2} \right]_+ ) \nonumber \\
\tilde f_4^d(u) &=&
\frac{1}{96 \pi^3} t E_1(t,s_0)  [(\delta(u) +
 4 {\bar u} - \left[ \frac{1}{u^2} \right]_+) \nonumber \\
\\
\label{DIM4u}
\tilde f_6^u(u) &=& \frac{4}{3} \delta(u) \nonumber \\
\tilde f_6^d(u) &=& \frac{2}{3} \delta({\bar u}) \\
\label{DIM6u}
\tilde f_8^u(u) &=& -\frac{4}{9} \delta(u) t^{-1} \nonumber \\
\tilde f_8^d(u) &=& - t^{-1} (\frac{2}{9} \delta({\bar u}) +
\frac{19}{54} \delta^\prime({\bar u}) )
\\ \nonumber
\label{DIM8u}
\end{eqnarray}
where for any test function f(u)
\begin{equation}
\int_0^1 du \left[ \frac{1}{u^2} \right]_+ f(u) \equiv
\int_0^1 du \frac{1}{u^2} (f(u) - f(0)- u f^\prime(0))\, .
\label{2plus}
\end{equation}
It is easy to see that in high orders of the OPE the series of power
corrections to the sum rules in coordinate and in momentum space will have
the following typical behaviour
\begin{eqnarray}
 \frac{\langle A\rangle^n}{t^n n!} z^n &\rightarrow&
\frac{\langle A\rangle^n}{t^n n!}  \delta^{(n)}(u)
\nonumber\\
  \frac{\langle A\rangle^n}{t^n n!} z^n \cos(z)&\rightarrow&
\frac{\langle A\rangle^n}{t^n n!}\delta^{(n)}(\bar u)
\end{eqnarray}
where $\langle A\rangle$ is the typical scale of vacuum fluctuations,
of order (several hundred MeV)$^2$, and the $n!$ suppression is due to
the Borel transformation.  For this reason the sum rules for
Ioffe-time distributions can be justified theoretically at small
values of $z$, while in the momentum space one faces a problem of the
summation of the series containing singular functions.

Let us now elaborate on this point.  The calculation of parton
distributions in coordinate space is essentially on the same theoretical
footing as the calculation of moments as matrix elements of local
operators. Mathematically, information about the moments is coded in
derivatives of the coordinate space distributions at $z=0$. Provided
OPE converges fast enough, the $z$-distribution is well defined and
can be calculated at sufficiently small $z$ by present nonperturbative
approaches to QCD. In the particular technique of QCD sum rules, the
results of calculations are usually assumed to be reliable provided
contributions of vacuum condensates are sufficiently small, say stay
within 30-40\% of the total. It turns out that for valence quark
distributions this criterium is satisfied for $z<3$. We shall see that
in practice the sum rule for u-quark distributions works in a larger
interval, and for d-quark distributions for a shorter interval; for
other approaches --- e.g. lattice calculations --- the limitations
can be different. Our point is that for sufficiently small $z$ one
does not need to invoke any additional assumptions. The contributions
of dimension 6 in (\ref{DIM6})  are respectable smooth functions at
small $z$ and must be taken into account, independent of their bad
behavior at large $z$ which produces $\delta$-functions after the
Fourier transform.

In momentum space, the calculation of parton distributions pointlike
in the Bjorken variable applies much more severe requirements to
non-perturbative techniques, and in practice requires additional
assumptions.  In particular, the approach of Ref. \cite{Iof1} assumes
that singular terms in the OPE do not affect calculation of parton
distributions at intermediate values of momentum fraction $u$, and
thus in this region all singular terms in the OPE can be neglected
altogether (see also \cite{many}).  This would be true if summation of
singular contributions produces a rather narrow smooth function with
the support either in $u\sim 0$ or in $u\sim 1$ regions.  Our task in
this paper is not to critisize this particular assumption, but rather
to make clear that assumptions of this kind are always necessary to
deal with parton distributions in momentum space, and thus provide an
additional input.

In fact, the assumption of Ref. \cite{Iof1} is non-trivial, and to our
opinion requires a better justification than given there.
Mathematically, the statement about calculability of coordinate-space
distributions at sufficiently small $z$ does not imply calculability
of momentum-space distributions at intermediate values of $u$.  We
find the neglect of singular contributions disturbing, since they are
100\% essential for calculation of the moments, see
Refs.\cite{kolesnichenko,BelBlok}. Since the calculation of the
parton distributions along the lines of Ref. \cite{Iof1} (and this
paper) is only justified as the analytic continuation from the
corresponding calculation of the moments \cite{bal88}, it is difficult
for us to imagine that important contributions to the moments of the
structure function will not show up in the distribution
itself. Physically, the assumption about the small smearing of
singular contributions implies existence of a certain second scale in
the hadrons, affecting the momentum distributions. We feel that a
further study of this question is necessary, to prove that smearing of
singular contributions does not affect the whole region of the Bjorken
variable.  Again, we repeat that advantage of coordinate space
formulation is that it avoids making any assumptions of this kind,
since singular contributions to not appear.

%%%%%%%%%%%Results%%%%%%%%%%%%%%%%%%%%%%%%%
Let us proceed to the description of our results. In the numerical
analysis  we
 use standard values of the parameters accepted in the QCD sum rules for the
nucleon i.e., $t \sim 1$ GeV$^2$ and $s_0 = (1.5$~GeV$)^2$, and the
following values for the condensates (the normalization point 1 GeV is
implied):
\begin{eqnarray}
\langle \bar q q \rangle &=& (250\,\mbox{\rm MeV})^3\,,
\nonumber\\
\langle(\alpha_s/\pi)G^2\rangle &=& 0.012\,\mbox{\rm GeV}^4\,,
\nonumber\\
\langle \bar q \sigma g G q \rangle &=& m_0^2\langle \bar q q\rangle;~~~
m_0^2 =0.64\,\mbox{\rm GeV}^2\, ,
\end{eqnarray}
which correspond to the standard ITEP values rescaled to the
normalization point $\mu^2 \sim M_N^2 \sim$ 1 GeV$^2$.

\begin{figure}[h]
\centerline{
\epsfysize=0.7\textheight
\epsfbox{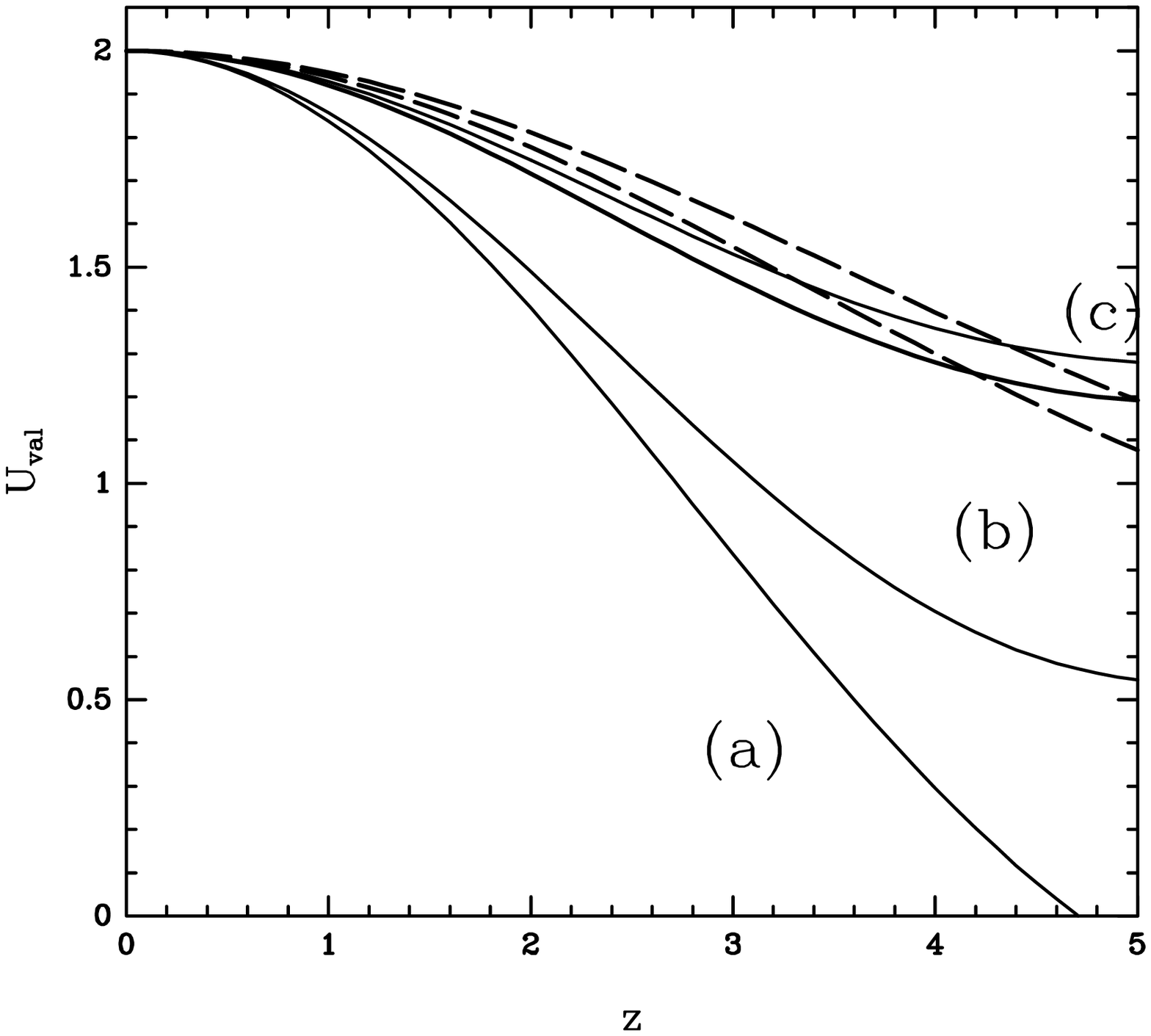}
}
\vspace{ 1.0in}
\caption[]{
QCD sum rules calculation of the valence u-quark Ioffe time
distribution function $U_{\rm val}(z,\mu^2)$ at $\mu^2 \sim 1$
GeV$^2$. The thick solid line results from the OPE with operators with
dimension 0, 4, 6 and 8. Thick dashed lines correspond to the
leading-order QCD analysis of Ref.\cite{GRVLO} at the scales  $\mu^2 =
0.5$ and 1 GeV$^2$, respectively. Solid line marked (a) is the
perturbative contribution to the sum rule.  Lines (b)
and (c) describe respectively the sum rules with operators of
dimension 4 and 6 taken into account. Note that VEV of dimension 8 gives
already a small contribution.
}\label{fig:6a}

\end{figure}

 The QCD sum rule prediction for the valence u-quark Ioffe time
distribution in the proton $U_{\rm val}(z,\mu^2 \sim 1$~GeV$^2)$ is
shown as the thick solid line in Fig.~\ref{fig:6a} and compared with
an ``experimental'' distribution. The latter has been obtained from
the leading-order parametrisation of Gluck, Reya and Vogt \cite{GRVLO}
normalized at 0.5 GeV$^2 \le \mu^2 \le$ 1 GeV$^2$.  We find
remarkable agreement up to rather large values $z \le 4$,
corresponding to longitudinal distances in the proton rest frame of
nearly 2 fm!  When this QCD sum rule result is augmented by the
assumption that $U_{\rm val}(z)$ is a sufficiently smooth function,
and combined with large--z behaviour implied by the Regge theory ( see
Eq.(\ref{AsEx})) it allows for complete reconstruction of $U_{\rm
val}(z)$ and therefore of the distribution function. Lines marked as
(a), (b) and (c) illustrate the relative importance of different
contributions to the sum rule, and are obtained keeping in
(\ref{AOPE}), (\ref{two-point}) the perturbative terms only (a),
adding the gluon condensate contribution (b) and adding in addition
also the $\langle\bar q q\rangle^2$ terms (c).  Note that the $U_{\rm
val}(z)$ distribution decreases at large $z$ more slowly than the
perturbative prediction, which is mainly due to bilocal corrections
arising from large distances in the t-channel. The latter can be
calculated as contact terms.  We remind that these terms are discarded
altogether in the approach of Ref. \cite{Iof1}.

\begin{figure}[h]
\centerline{
\epsfysize=0.7\textheight
\epsfbox{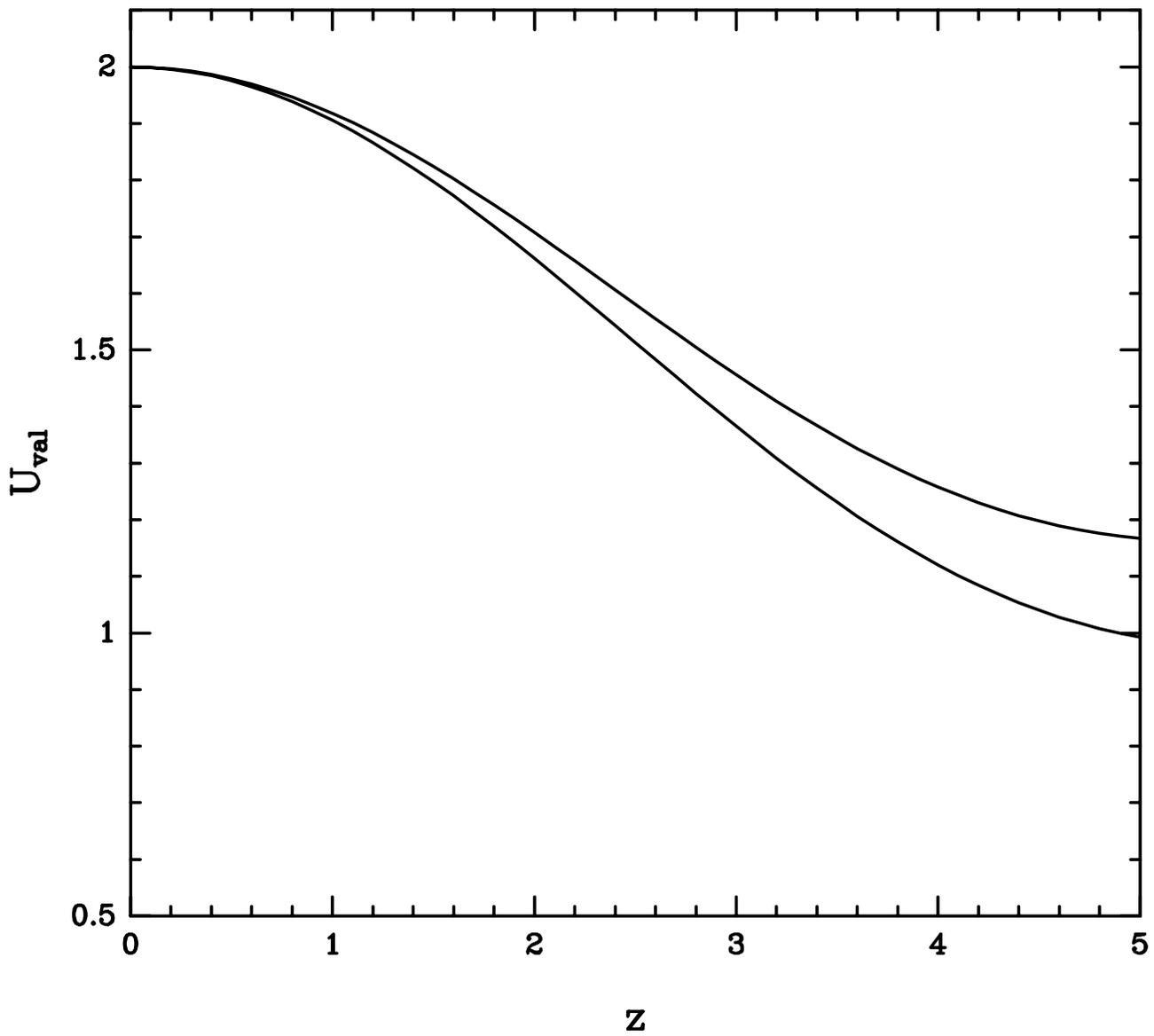}
}
\vspace{ 1.0in}
\caption[]{
Stability of the sum rule for $U_{\rm val}(z)$ against variation of the Borel
parameter t. The upper curve corresponds to t = 1 GeV$^2$, the lower
to t = 1.5 GeV$^2$.
}\label{fig:6b}

\end{figure}

 Figure \ref{fig:6b} shows stability of our prediction when the Borel
parameter t is varied between 1 GeV$^2$ (upper curve) and 1.5 GeV$^2$
(lower curve). One may conclude that in the region $z \le 4$ the
valence u-quark sum rule converges very fast and it is
numerically stable.

A sum rule similar to the one given by Eq.~(\ref{AOPE}) can be written
for a non-zero value of the momentum transfer $q^2$ in the t-channel,
allowing to study the radius of the valence quark distributions
(cf.\cite{BGMS}).  We have checked that the radius of the valence
u-quark distribution obtained from the sum rule in (\ref{AOPE}) is
close to the measured electromagnetic radius of the proton, which
is encouraging.

\begin{figure}[h]
\centerline{
\epsfysize=0.7\textheight
\epsfbox{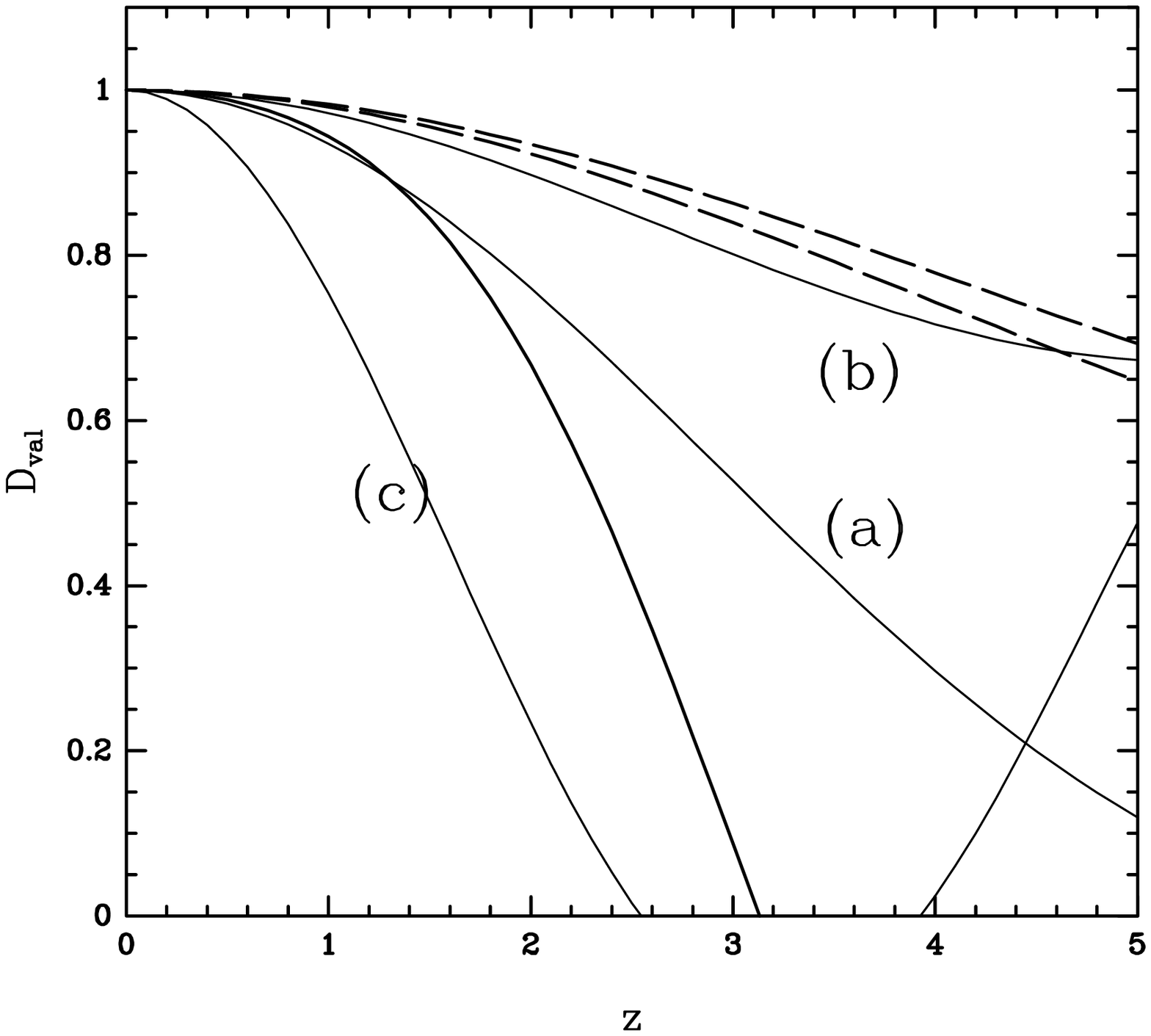}
}
\vspace{ 1.0in}
\caption[]{
QCD sum rules calculation of the valence d-quark Ioffe time
distribution function $D_{\rm val}(z,\mu^2)$ at $\mu^2 \sim 1$
GeV$^2$. The thick solid line results from the OPE with operators of
dimension 0, 4, 6 and 8. Thick dashed lines correspond to the
leading-order QCD analysis of Ref.\cite{GRVLO} at the scales $\mu^2 =
0.5$ and 1 GeV$^2$, respectively. Solid line marked (a) is the
perturbative contribution to the sum rule.  Lines (b)
and (c) describe respectively the sum rules with condensates of
dimension 4 and 6 taken into account.
}\label{fig:7}

\end{figure}

The situation is not so good, unfortunately, for the valence d-quark
distribution, see Fig.~\ref{fig:7}. The sum rule prediction for
$D_{\rm val}(z)$, shown as the thick solid line, is rather far from
the leading-order GRV \cite{GRVLO} parametrisation, and is much more
short-range. As in Fig.~\ref{fig:6a}, we also show contributions of
various terms in the OPE to the final result. It is seen that taking
into account the gluon condensate contribution (b) improves the
prediction considerably compared to the perturbative result (a), but
this tendency is destroyed by the contributions of dimension 6 (c) and
dimension 8. In the case of the d-quark distribution the bilocal
corrections corresponding to the contact terms are absent, and the
problem arises because of the diagram in which all nucleon momentum is
carried by a single quark, see Fig.~\ref{fig:4} (d), which contributes
a term $\sim \cos(z)$ (\ref{DIM6}).  In the momentum space this
contribution is proportional to $\delta(1-u)$, where u is a fraction
of longitudinal momentum carried by the quark, see (\ref{DIM6u}).
Another reason is the absence of the bilocal power correcton of dimension
6, contributing a term $\sim \delta(u)$ in (\ref{DIM6u}), suggesting that
terms of higher dimension $\sim \delta(u), \delta'(u)\ldots$ can be
important.

It turns out that the d-quark distribution is very sensitive to the
numerical values of condensates of dimension 6 and 8 because of strong
cancellation between corresponding terms. If we took ${\bar q}q$ =
- (240 MeV)$^3$ and $m_0^2 = 0.8$ GeV$^2$, which correspond to the
normalization point $\mu^2 = 0.5$ GeV$^2$, we would obtain perfect
agreement with the experimental analysis up to values of z of order of
1, where the sum rules prediction would abruptly turn down. Such a
strong normalization-point dependendence makes our prediction for d
quarks less reliable.

A favourable structure of the OPE for the u-quark distribution and the
complications for d-quarks have, presumably, no physical relevance,
and are due to the particular structure of the interpolating current
(\ref{Eta}).  This choice is standard, but, as it follows from our
analysis, not very convenient for the study of the d-quark
distributions, since it implies that the correct behaviour of this
distribution is due to higher order power corrections.

A more detailed analysis of this problem goes beyond the scope of this
paper. One could try a different interpolating current to improve the
results for the d-quark distributions, or calculate radiative
corrections to the sum rule, which generally tend to soften the parton
distributions (i.e., make them more extended in the Ioffe time) and
are expected to be especially important for d-quarks, see \cite{Iof1}.

However, it is worthwhile to demonstrate, at least semiquantitatively,
that the higher order contributions indeed tend to smoothen the
$\sim\delta(1-u)$ contribution of dimension 6 in the d-quark sum rule,
and are potentially able to bring it to the agreement with the data.
To this end we use the concept of non-local condensates, introduced in
\cite{Rad}, which allows to consider the effects of the final
correlation length in the QCD vacuum, the property that is missing in
the local operator product expansion.

Note that the contribution of the diagram in Fig.~\ref{fig:4} (d) is
essentially proportional to the vacuum expectation value of the
nonlocal operator $(u^{aT}(x)C\gamma_\xi u^b(x))({\bar
u}^a(0)C\gamma_\xi {\bar u}^{bT}(0))$ which produces the expansion
\begin{equation}
\langle (u^{aT}(x)C\gamma_\xi u^b(x))
({\bar u}^a(0)C\gamma_\xi {\bar u}^{bT}(0))\rangle
= -\frac{2}{3} \langle \bar q q \rangle^2
[1 +\frac{1}{8} m_0^2 x^2+\ldots]
\end{equation}
where we have assumed the factorization to evaluate the coefficients.
It is the expansion into the sum of local operators that generates the
series of power corrections proportional to derivatives of
$\delta(1-u)$ in the sum rule for the momentum fraction distributions.
As noted in \cite{Rad}, this expansion misses an important property of
the correlation functions in Euclidian space, which is in existence of
the final correlation length in the physical vacuum.  To illustrate
this point, let us consider the exponential parametrization
\begin{equation}
\langle (u^{aT}(x)C\gamma_\xi u^b(x))
({\bar u}^a(0)C\gamma_\xi {\bar u}^{bT}(0))\rangle =
 -\frac{2}{3} \langle {\bar q} q\rangle^2 \int_0^\infty d\nu\,
e^{\frac{x^2}{4}\nu} f(\nu)
\label{NLC}
 \end{equation}
Moments of the function $f(\nu)$ are determined by vacuum expectation
values of local operators. The first few of them in the factorization
approximation are fixed to be
\begin{eqnarray}
\int_0^\infty d\nu f(\nu) &=& 1 \nonumber \\
\int_0^\infty d\nu\, \nu f(\nu) &=& \frac{1}{2}
{{\langle{\bar q}g \sigma G q\rangle }
\over{\langle{\bar q}q\rangle}} \equiv \frac{1}{2} m_0^2\,.
\label{NLC1}
\end{eqnarray}
On the other hand, one generally expects that the correlation functions
in QCD decrease exponentially in Euclidian space, suggesting that
\begin{equation}
\langle (u^{aT}(x)C\gamma_\xi u^b(x))
({\bar u}^a(0)C\gamma_\xi {\bar u}^{bT}(0))\rangle
\sim \exp[-M_D\cdot\sqrt{-x^2}]
\end{equation}
at large $x^2\to -\infty $, where $M_D$ is the correlation length which,
loosely speaking, may be associated with the diquark mass.
It is easy to see that this behaviour corresponds to the asymptotics
\begin{equation}
f(\nu) \sim e^{-{\rm M_D^2}/\nu}
\end{equation}
at small $\nu$. Note that the expansion into the sum of local
operators corresponds to the expansion of $f(\nu)$ in delta-functions
at $\nu=0$.

The effect of using the nonlocal condensate in the sum rules is easy
to evaluate. In coordinate space, insertion of (\ref{NLC}) amounts to the
substitution of the coefficient function $f_6^d$ in (\ref{DIM6})
by
\begin{equation}
f_{NLC}^d(z,t) = \frac{2}{3} \langle{\bar q}q\rangle^2
t \int_0^1 du\, u f({\bar u}t)\,\cos(u z)\,.
\label{NLC2}
\end{equation}
In momentum space this replacement is simply
\begin{equation}
\tilde f_6^d(u) =\frac{2}{3}\delta(\bar u)~~ \rightarrow~~
\tilde f_{NLC}^d(u,t) =\frac{2}{3}t u f(\bar u t)
\label{NLC3}
\end{equation}
For numerical estimates we choose a simple model
\begin{equation}
f(\nu) = {(M_D^2)^{a-2}\over{\Gamma(a-2)}} \nu^{1-a} e^{-{\rm M_D^2}/\nu}
\end{equation}
with two parameters $M_D$ and $a$. Equation (\ref{NLC1}) leads to the
constraint
\begin{equation}
a - 3 = 2 \frac{M_D^2}{m_0^2}
\end{equation}
so that we only need to specify the correlation length. Its value has
a direct physical meaning and is related to the difference between
masses of heavy baryons, containing $b$ quark and the $uu$ pair, and
the mass the $b$ quark.\footnote{ We do not discuss this issue in
detail, referring to a well-known relation between the asymptotics of
the quark propagator at large distances in Euclidian space, and the
difference between heavy meson and quark masses in the heavy quark
limit \cite{Rad2}.  This difference is usually denoted by $\bar
\Lambda$ and is one of the main observables in the heavy quark
effective theory. For baryons the situation is quite similar.  The
range of values of $M_D$ used here corresponds to the estimates found
in the literature.}

\begin{figure}[h]
\centerline{
\epsfysize=0.7\textheight
\epsfbox{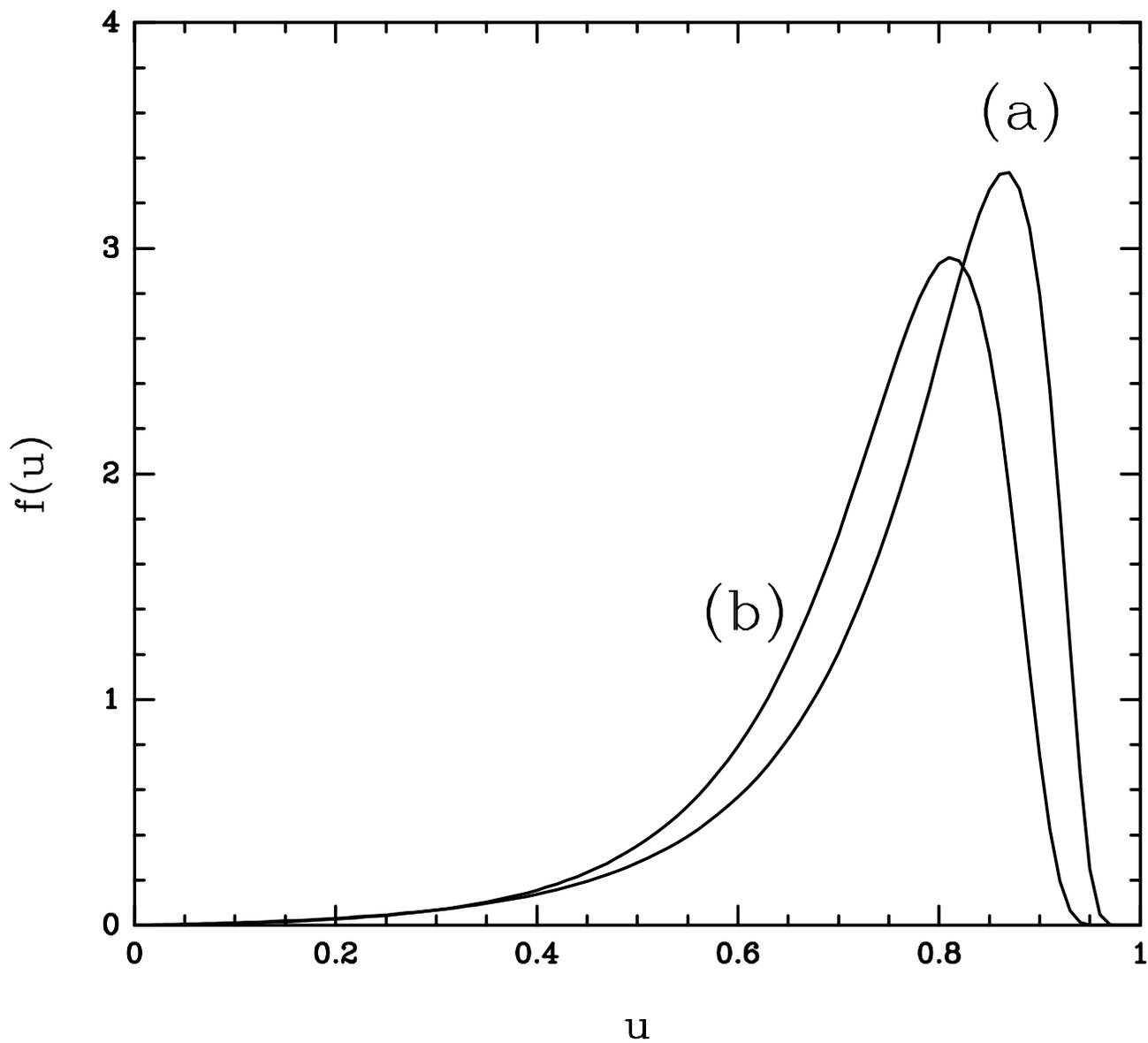}
}
\vspace{ 1.0in}
\caption[]{
The nonlocal condensate contribution to the d-quark sum rule, see
Eq.(\protect{\ref{NLC3}}). Lines marked as (a) and (b) correspond to
the diquark mass parameters $M_D = 0.7$ GeV and 1.0 GeV,
respectively. Borel parameter t = 1 GeV$^2$.  }\label{fig:8}

\end{figure}

In Fig.~\ref{fig:8} we show the r.h.s. of (\ref{NLC3}) as a function
of the momentum fraction $u$ for the Borel parameter $t=1$ GeV$^2$ and
for two choices of $M_D=700$ MeV and 1 GeV. We see that the main
effect of the finite correlation length is to push nonperturbative
effects $\sim \langle \bar q q\rangle^2$ away from the region $u\sim
1$.  This may seem contrary to the physical intuition, but not
necessarily so, because the effects of the quark condensate are
qualitatively similar to the introduction of the constituent mass.  We
remark that this picture contradicts the expectations of
Ref. \cite{Iof1}: Summation of singular contributions not only does
not produce a narrow function with a support concentrated at $u\to 1$,
but, on the contrary, the nonperturbative contributions die away at
$u\to 1$ faster than any power of $1-u$. We stress that this is a
direct consequence of of the final correlation length in the QCD
vacuum. Our model estimates presented in Fig.~\ref{fig:8} show that
the resulting contributions are important at least up to $u\sim 0.6$.

\begin{figure}[h]
\centerline{
\epsfysize=0.7\textheight
\epsfbox{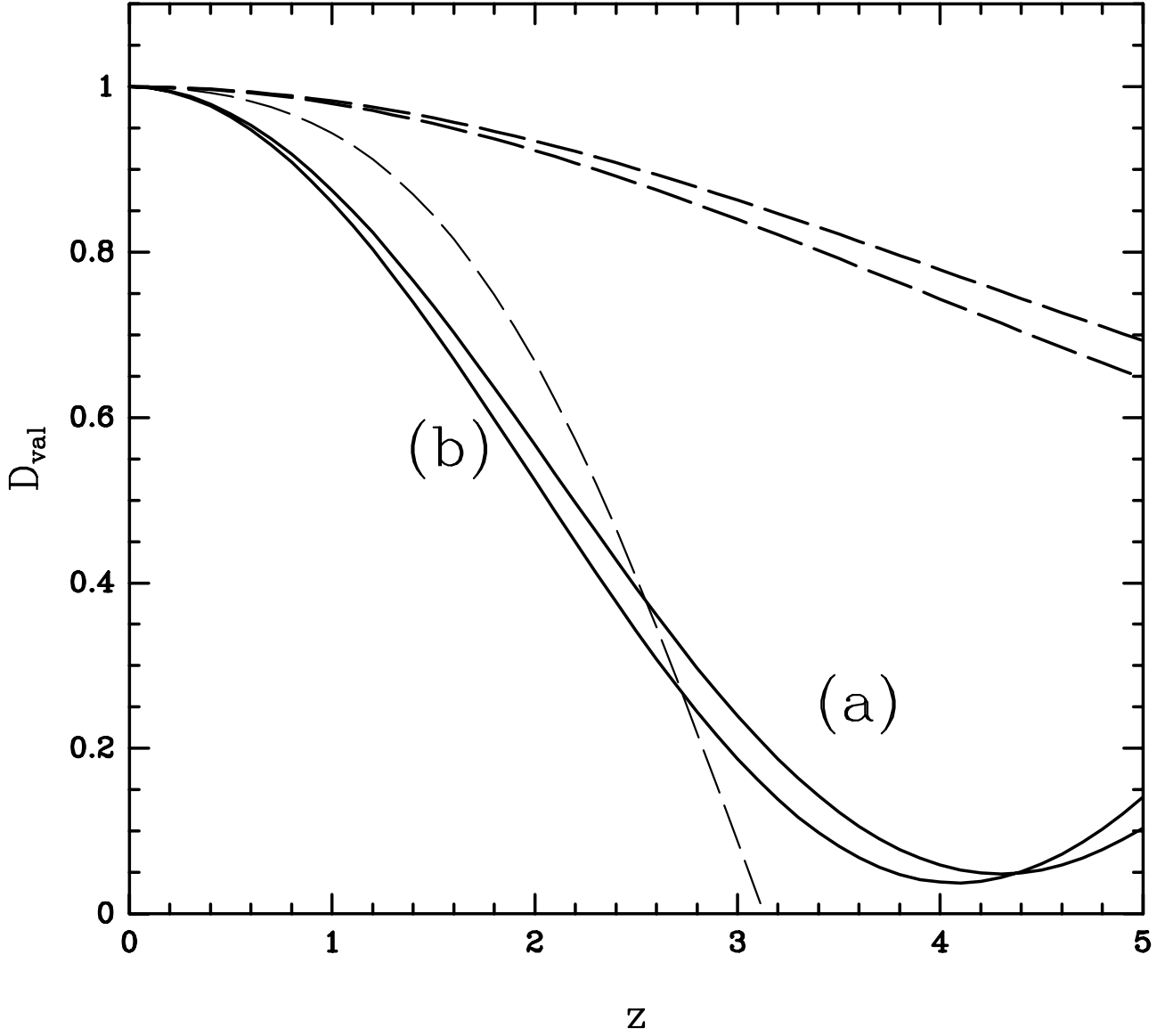}
}
\vspace{ 1.0in}
\caption[]{
QCD sum rules calculation of the valence d-quark Ioffe time
distribution function $D_{\rm val}(z)$ (solid lines). VEV of dimension
6 and 8 has been replaced by the phenomenological model of non-local
four-quark condensate, Eq.(\protect{\ref{NLC}}). Labels (a) and (b)
refer to the diquark mass parameters $M_D = 0.7$ GeV and 1.0 GeV,
respectively. Thick dashed lines correspond to the leading-order QCD
analysis of Ref.\cite{GRVLO} normalized at $\mu^2 = 0.5$ and 1
GeV$^2$, respectively. The dashed line shows results of the standard
sum rule (\protect{\ref{AOPE}}).  }\label{fig:9}

\end{figure}

Numerical results for the valence d quark distribution $D_v(z)$ with
the nonlocal condensate are shown in Fig.~\ref{fig:9}. We see that the
situation improved somewhat at $z>2$, although this type of
contributions alone is not able to restore the agreement to the data.
A better agreement can be obtained by choosing a larger value of the
mixed condensate parameter $m_0^2$, but this possibility is not very
attractive.  We expect, however, that the results will be substantially
improved by taking into account radiative corrections.
An inspection shows that the sum rule for d quark distributions can also
be saved by a large bilocal power correction, contributing a term $\sim z^2$
in coordinate space (alias $\sim \delta''(u)$ in momentum space).
These corrections
are difficult to evaluate, however, and we do not attempt this task in the
present paper. An experience of QCD sum rule calculations generally
suggests that if there are indications that the sum rule is affected by
contributions of high order in the OPE, it is advisable to use different
interpolating currents for the participating hadrons.

\section{Summary and Conclusions}

%%%%%%CONCLUSIONS%%%%%%%%%

We suggest to use Ioffe time distributions -- the distributions of
invariant longitudinal distances z essential in a deep-inelastic
scattering process -- as a suitable alternative to the conventional
description in terms of momentum fraction parton distributions. The
advantage of this formulation is that contributions of large and small
longitudinal distances that correspond in fact to different physics,
become in this approach separated.  The large-distances of
distributions in coordinate space are governed by the Regge theory,
and can be taken as an input, while the calculation at moderate
distances can be within reach of current nonperturbative approaches,
e.g.  the lattice QCD.  We illustrate these advantages using the QCD
sum rule technique to calculate valence quark distributions, and
compare it to the corresponding calculations in the momentum space.
Although the sum rules derived in this paper are rather preliminary
and can be improved significantly by taking into account further
corrections, we obtain a very good description of the u quark
distributions. The results for the d quark distributions are worse, the
reason is an unfavourable structure of the operator product
expansion series in this case, where higher order terms are important.

We believe that our results can be improved by making a
state-of-the-art QCD sum rule analysis, or with different
techniques. In particular, we expect that Ioffe time distributions may
be feasible for lattice calculations and for instanton models of the
QCD vacuum of the type suggested in Ref. \cite{shuryak}. It would be
most interesting to constrain in this way the polarized gluon
distribution which is poorly known.

\begin{center}
{\bf Acknowledgments}
\end{center}
This work has been supported by KBN grant 2~P302~143~06. We are
grateful to the referee for pointing out the error in the earlier
version of this manuscript and for bringing ref. \cite{Tera} to our
attention.

%%%%%%%APPENDIX%%%%%%%%%%%
\appendix

\renewcommand{\theequation}{\Alph{section}.\arabic{equation}}
\section*{ Appendix A}
\setcounter{equation}{0}
\setcounter{section}{1}

In this Appendix we derive a Ward identity for the correlation function
(\ref{PiQ}) useful in practical calculations. In addition, it makes
normalization properties of valence quark distributions explicit.

In our special kinematics vectors
$n_\mu$ and $q_\mu$ are proportional:
\begin{equation}
n_\mu = \frac{n \cdot x}{q \cdot x} q_\mu.
\label{DpQ}
\end{equation}
 Replacing $\gamma_\mu n^\mu$ in the definition (\ref{OurOp})
of ${\hat O}^i$ by $\frac{n \cdot x}{q \cdot x} \gamma_\mu
q^\mu$ and integrating by parts over $d^4y$ in (\ref{PiQ})
we get:
\begin{eqnarray}
\Pi^i & = &  - i  \int  d^4x\, \exp {(i p\cdot x)}
 \int d^4y\, \exp {(i q \cdot y)}
\frac{n \cdot x}{q \cdot x} \nonumber \\
& ~ & \frac{\partial}{\partial y_\mu}
\langle \Omega \mid T\{\eta(x) {\bar \Psi}^i(y+\frac{\Delta}{2}) \gamma_\mu
[y+\frac{\Delta}{2};y-\frac{\Delta}{2}]
\Psi^i(y-\frac{\Delta}{2}) {\bar \eta}(0) \} \mid \Omega \rangle
\nonumber \\
&  &{}+ (\Delta \to - \Delta) \; .
\label{W1}
\end{eqnarray}
The next step is to evaluate the derivative ${\partial}/{\partial
y_\mu}$ explicitly. One gets
\begin{eqnarray}
& ~ & \frac{\partial}{\partial y_\mu}
{\bar \Psi}^i(y+\frac{\Delta}{2}) \gamma_\mu
[y+\frac{\Delta}{2};y-\frac{\Delta}{2}]
\Psi^i(y-\frac{\Delta}{2}) = \nonumber \\
& ~ & {\hat D} {\bar \Psi}^i(y+
\frac{\Delta}{2})[y+\frac{\Delta}{2};y-\frac{\Delta}{2}]
\Psi^i(y-\frac{\Delta}{2}) +
{\bar \Psi}^i(y+\frac{\Delta}{2})
[y+\frac{\Delta}{2};y-\frac{\Delta}{2}]
{\hat D}\Psi^i(y-\frac{\Delta}{2}) \nonumber \\
& ~ & - \frac{i g}{2} \int_{-1}^1 d\xi
 {\bar \Psi}^i(y+\frac{\Delta}{2})
 [y+\frac{\Delta}{2};y+\xi\frac{\Delta}{2}]
\gamma^\alpha \Delta^\beta G_{\alpha\beta}(y+\xi\frac{\Delta}{2})
[y+\xi\frac{\Delta}{2};y-\frac{\Delta}{2}]
\Psi(y-\frac{\Delta}{2}), \nonumber \\
\label{W2}
\end{eqnarray}
where $G_{\alpha\beta}$ is the gluon field strength.

\begin{figure}[h]
\centerline{
\epsfysize=\textheight
\epsfbox{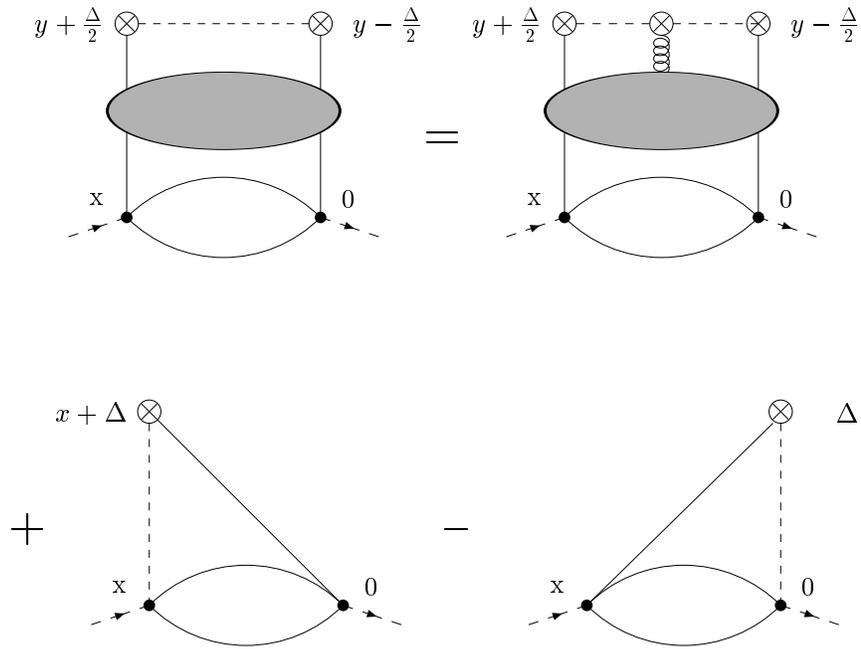}
}
\vspace{ - 1.0in}
\caption[]{
Graphical illustration of the Ward identity,
Eq.(\protect{\ref{W4}}). Dashed lines denote path-ordered
exponentials.  }\label{fig:10}

\end{figure}

Inserting the right-hand-side of (\ref{W2}) into (\ref{W1}) we note
that terms containing ${\hat D}{\bar \Psi}$ and ${\hat D} \Psi$ lead
to contact terms.  To this end it is convenient to have in mind the
functional integral representation of the correlation function in
(\ref{W1}), and use the identity
\begin{equation}
\exp\left( i \int{\cal L}\, d^4 x \right){\hat D} \Psi(y) =
- \frac{\delta}{\delta {\bar \Psi}(y)}\exp\left( i \int{\cal L}\, d^4 x
\right)
\label{W3}
\end{equation}
where ${\cal L}$ is the QCD Lagrangian.
Making an integration by parts in the functional integral
we obtain the delta
function $\delta(x-y + \frac{\Delta}{2})$ which allows to perform the
integration over variable $y$ explicitly. The net result can be
written as a Ward identity:
\begin{eqnarray}
\Pi^i & = & \frac{1}{2}\int d^4x\, \exp{(i p \cdot x)} \int dy\,
\exp{(i q \cdot y)}
\,\frac{n \cdot x}{q \cdot x} \int_{-1}^1 d\xi\ \langle \Omega \mid
T\Big\{ \eta(x) {\bar \Psi}^i(y+\frac{\Delta}{2})
\nonumber\\
&&{}\times
[y+\frac{\Delta}{2};y+\xi\frac{\Delta}{2}]
 g \gamma^\alpha \Delta^\beta G_{\alpha\beta}
(y+\xi\frac{\Delta}{2})
[y+\xi\frac{\Delta}{2}; y-\frac{\Delta}{2}]
 \Psi^i(y-\frac{\Delta}{2}) {\bar \eta}(0)\Big\} \mid
\Omega \rangle  \nonumber \\
&&{}+ N^i i \int d^4x\, \exp{(i p \cdot x)}
 \frac{n \cdot x}{q \cdot x}
\left[ \exp{(i q \cdot x)}
\langle \Omega \mid T (\eta^i(x;- \Delta) {\bar \eta(0)})
\mid \Omega \rangle \right. \nonumber \\
& &{}- \left. \langle \Omega \mid T (\eta(x) {\bar \eta}^i(0;\Delta))
\mid \Omega \rangle\right] + {}(\Delta \to - \Delta)
\label{W4}
\end{eqnarray}
where $N^i =$ 1 and 2 for d- and u-quarks, respectively. The {\em
nonlocal} currents $\eta^i(x;-\Delta)$ are defined as
\begin{eqnarray}
\eta^d(x;-\Delta) = \epsilon^{abc} u^a(x)^T C\gamma_\mu u^b(x)
\gamma_5 \gamma^\mu [x;x-\Delta]^{cf} d^f(x-\Delta)
\nonumber \\
\eta^u(x;-\Delta) = \epsilon^{abc} u^{fT}(x-\Delta)C \gamma_\mu
[x-\Delta;x]^{fa} u^b(x) \gamma_5 \gamma^\mu d^c(x)
\label{W50}
\end{eqnarray}
%%%%%%%%%%%%%%%%%%%%
The outcome of Eq. (\ref{W4}) is that the complicated correlation
function in (\ref{PiQ}) is written as a sum of several simpler terms.
The last two terms contain one integration less compared to the
original expression, and therefore they cannot generate bilocal power
corrections.  The first term contains explicitly a gluon field. Thus,
the OPE for this term starts with higher orders in the coupling and
(or) the dimension of the corresponding operators.  By an explicit
comparison of the OPE applied to the correlation function (\ref{PiQ})
and to its equivalent form in (\ref{W4}) one can make sure that
several important BPC's are transformed in this way to the LPC's
related to vacuum expectation values of local operators, see
Fig.~\ref{fig:10} for the illustration. In fact, the last two terms
in (\ref{W4}) presumably collect {\em all} BPC's which can be reduced
to vacuum condensates by low-energy theorems.  Experience of
practical calculations in the sum rules shows that these terms are
typically the most important ones numerically. In this paper we do
not take into account additional BPC's from the expansion of the term
with an extra gluon in (\ref{W4}).

In the limit  $\Delta\to 0$  the last term disappears and we are left
with the correlation function
\begin{equation}
{ \Pi}^i  =  N^i i \int d^4x \exp{(ip\cdot x)} \frac{n
\cdot x}{q \cdot x}
\left( \exp{(i q \cdot x)} - 1 \right) \langle
\Omega \mid T \left( \eta(x)
{\bar \eta}(0)\right)\mid \Omega \rangle \; .
\label{WW6}
\end{equation}
Equation (\ref{WW6}) can further be simplified writing
\begin{equation}
\left( \exp{(i q \cdot x)} - 1 \right) = i q \cdot x
\int_0^1 dv\, \exp{(i v q \cdot x)}\; ,
\label{WW6a}
\end{equation}
so that we finally obtain
\begin{eqnarray}
{\Pi}^i & = & N^i \int_0^1 dv\, n_\mu
\frac{\partial}{\partial {\tilde p}_\mu}
i \int d^4 x \exp{(i {\tilde p}\cdot x)}
\langle \Omega \mid T\left( \eta(x)
{\bar \eta}(0) \right) \mid \Omega \rangle \nonumber \\
& = & N^i \int_0^1 dv\, n_\mu
\frac{\partial}{\partial {\tilde p}_\mu}
\Pi^{(2)}({\tilde p}) ,
\label{W6}
\end{eqnarray}
where ${\tilde p} = p + v q$, arriving at the equation in (\ref{W5}).

As an illustration of the use of the identity in (\ref{W4}) in
practical calculation, let us consider the leading perturbative
contribution (PT) to the correlation function (\ref{PiQ1}) for u quarks,
shown in Fig.~\ref{fig:4}(a).  First we note that owing to the
explicit presence of the coupling in the gluon field, the first term
in (\ref{W4}) can to this accuracy be neglected.  Since
\begin{equation}
\langle \Omega \mid T (\eta^u(x;- \Delta) {\bar \eta(0)})
\mid \Omega \rangle_{PT} =
\langle \Omega \mid T (\eta(x) {\bar \eta}^u(0;\Delta))
\mid \Omega \rangle_{PT} \; ,
\label{W7}
\end{equation}
the remaining contribution can be rewritten as:
\begin{eqnarray}
\Pi^u_{PT} & = & N^u i \int d^4x\, \exp{(i p \cdot x)}
 \frac{n \cdot x}{q \cdot x}
\left( \exp{(i q \cdot x)} - 1\right)
\langle \Omega \mid T (\eta^u(x;- \Delta) {\bar \eta(0)})
\mid \Omega \rangle_{PT} \, . \nonumber \\
\label{W8}
\end{eqnarray}
and further using (\ref{WW6}), as
\begin{equation}
\Pi^u_{PT}  =  N^u i \int_0^1 dv \int d^4x\, \exp{(i {\tilde p} \cdot x)}
(n \cdot x)
\langle \Omega \mid T (\eta^u(x;- \Delta) {\bar \eta(0)})
\mid \Omega \rangle_{PT} \, .
\label{W9}
\end{equation}
A straightforward calculation of (\ref{W9}) using dimensional
regularisation produces
\begin{equation}
\frac{1}{4} {\rm Tr} {\not\! n \,}
\Pi^u = \frac{1}{4 \pi^4} (p \cdot n)^2
\int_0^1 dv \frac{\Gamma(3-d)}{[- {\bar v} p_1^2 - v p_2^2]^{3-d}}
\int_0^1 du\; (9 u {\bar u}^2 + {\bar u}^3) \cos{uz} \; ,
\label{W10}
\end{equation}
Comparing with equation (\ref{PiQ1}) and performing the Borel
transformation according to (\ref{Borel0}) we finally arrive at the
u-quark coefficient function in equation (\ref{PT}).

\vfill\eject
\section*{Figure captions}
\begin{description}
\item[Fig.~1]
Ioffe time distributions for partons at the scale $\mu^2 = 4$
GeV$^2$. The solid line denotes Gl\"uck, Reya, Vogt (GRV)
\cite{GRVNLO} set of parametrizations, the dashed line denotes CTEQ
\cite{CTEQ} parametrization. $U_{\rm val}(z)$ and $D_{\rm val}(z)$ are
the valence quarks distributions, $Q_U(z)$ and $Q_D(z)$ are the C-even
up and down quark distributions, and $G(z)$ is the gluon distribution.
\item[Fig.~2]
Model of the Ioffe time valence-type distribution corresponding to the
simple ansatz $q_V(u) = N u^{\alpha} (1-u)^\beta$ with $\alpha = - 0.5$
and $\beta = 3$. The normalization is such that $\int_0^1 \, du
q_V(u) = 1$. The dashed line shows the asymptotic expansion
(\ref{AsEx}). Note that it matches almost perfectly the true behaviour
for $z \ge 6$.
\item[Fig.~3]
Model of the Ioffe time distributions for polarized gluon density
$\Delta g(u) = N_G u^\alpha (1-u)^\beta$ with $\alpha = 0$ and $\beta
= 4$. The normalization constant $N_G$ is chosen in such a way that
the gluon polarisation $\Delta g = 0.5$. The two short-dashed curves
were obtained by taking $\beta$ equal to 3.5 and 4.5 respectively, and
keeping $\alpha=0$ and $\Delta g=0.5$ fixed. The long-dashed curve is
the asymptotic expansion (\ref{AsEx}).
\item[Fig.~4]
Evolution of valence $U_{\rm val}(z,Q^2)$ and $D_{\rm val}(z,Q^2)$,
and gluon $G(z,Q^2)$ Ioffe time distributions. GRV \cite{GRVNLO}
parametrisation has been used. The solid line corresponds to $Q^2 = 4$
GeV$^2$, the dashed line to $Q^2 = 20$ GeV$^2$, respectively.
\item[Fig.~5]
Evolution of up and down quark C-even $Q_U(z,Q^2)$ and $Q_D(z,Q^2)$
Ioffe time distributions. GRV \cite{GRVNLO} parametrisation has been
used. The solid line corresponds to $Q^2 = 4$ GeV$^2$, the dashed line
to $Q^2 = 20$ GeV$^2$, respectively.
\item[Fig.~6]
Typical diagrams contributing to the OPE of the correlator (\ref{PiQ}).
\item[Fig.~7.]
Generic form of a bilocal correction in the OPE of the
correlator (\ref{PiQ}).
\item[Fig.~8]
QCD sum rules calculation of the valence u-quark Ioffe time
distribution function $U_{\rm val}(z,\mu^2)$ at $\mu^2 \sim 1$
GeV$^2$. The thick solid line results from the OPE with operators with
dimension 0, 4, 6 and 8. Thick dashed lines correspond to the
leading-order QCD analysis of Ref.\cite{GRVLO} at the scales  $\mu^2 =
0.5$ and 1 GeV$^2$, respectively. Solid line marked (a) is the
perturbative contribution to the sum rule.  Lines (b)
and (c) describe respectively the sum rules with operators of
dimension 4 and 6 taken into account. Note that VEV of dimension 8 gives
already a small contribution.
\item[Fig.~9]
Stability of the sum rule for $U_{\rm val}(z)$ against variation of the Borel
parameter t. The upper curve corresponds to t = 1 GeV$^2$, the lower
to t = 1.5 GeV$^2$.
\item[Fig.~10]
QCD sum rules calculation of the valence d-quark Ioffe time
distribution function $D_{\rm val}(z,\mu^2)$ at $\mu^2 \sim 1$
GeV$^2$. The thick solid line results from the OPE with operators of
dimension 0, 4, 6 and 8. Thick dashed lines correspond to the
leading-order QCD analysis of Ref.\cite{GRVLO} at the scales $\mu^2 =
0.5$ and 1 GeV$^2$, respectively. Solid line marked (a) is the
perturbative contribution to the sum rule.  Lines (b)
and (c) describe respectively the sum rules with condensates of
dimension 4 and 6 taken into account.
\item[Fig.~11]
The nonlocal condensate contribution to the d-quark sum rule, see
Eq.(\ref{NLC3}). Lines marked as (a) and (b) correspond to the diquark mass
parameters $M_D = 0.7$ GeV and 1.0 GeV, respectively. Borel parameter
t = 1 GeV$^2$.
\item[Fig.~12]
QCD sum rules calculation of the valence d-quark Ioffe time
distribution function $D_{\rm val}(z)$ (solid lines). VEV of dimension
6 and 8 has been replaced by the phenomenological model of non-local
four-quark condensate, Eq.(\ref{NLC}). Labels (a) and (b) refer to
the diquark mass parameters $M_D = 0.7$ GeV and 1.0 GeV,
respectively. Thick dashed lines correspond to the leading-order QCD
analysis of Ref.\cite{GRVLO} normalized at $\mu^2 = 0.5$ and 1
GeV$^2$, respectively. The dashed line shows results of the standard
sum rule (\ref{AOPE}).
\item[Fig.~13]
Graphical illustration of the Ward identity, Eq.(\ref{W4}). Dashed lines
denote path-ordered exponentials.

\end{description}

\vfill
\eject

%
%TO RUN WITHOUT FIGURES, UNCOMMENT HERE
%\end{document}
%

\renewcommand{\textfraction}{0}
\clearpage

%\section*{Figures}

\clearpage

\clearpage

\clearpage

\clearpage

\clearpage

\clearpage

\clearpage

\clearpage

\clearpage

\clearpage

\clearpage

\clearpage

\clearpage

\end{document}